\documentclass[aps,prd,preprint,eqsecnum,tightenlines,floats,
groupedaddress,showpacs]{revtex4}
\usepackage{amssymb} 

\usepackage{graphicx}
\usepackage{subfigure}
\usepackage{dcolumn}

\begin{document}

\date{March 28, 2023}

\title{Low Frequency (100 - 600 MHz) Searches with Axion Cavity Haloscopes}

\author{S. Chakrabarty}
\author{J.~R.~Gleason}
\author{Y. Han}
\author{A. T. Hipp}
\author{M. Solano}
\author{P. Sikivie}
\author{N. S. Sullivan}
\author{D. B. Tanner}
\vskip -3mm
\affiliation{University of Florida, Gainesville, Florida 32611, USA}

\author{M. Goryachev}
\author{E. Hartman}
\author{B. T. McAllister}
\author{A. Quiskamp}
\author{C. Thomson}
\author{M. E. Tobar}
  \affiliation{University of Western Australia, Perth, Western Australia 
6009, Australia}

\author{M. H. Awida}
\author{A. S. Chou}
\author{M. Hollister}
\author{S. Knirck}
\author{A. Sonnenschein}
\author{W. Wester}
  \affiliation{Fermi National Accelerator Laboratory, Batavia, Illinois 
60510, USA}

\author{T. Braine}
\author{M. Guzzetti}
\author{C. Hanretty}
\author{G. Leum}
\author{L. J Rosenberg}
\author{G. Rybka}
\author{J. Sinnis}
  \affiliation{University of Washington, Seattle, Washington 98195, USA}

\author{John Clarke}
\author{I. Siddiqi}
  \affiliation{University of California, Berkeley, California 94720, USA}

\author{R. Khatiwada}
  \affiliation{Illinois Institute of Technology, Chicago, Illinois 60616, 
USA}
  \affiliation{Fermi National Accelerator Laboratory, Batavia, Illinois 
60510, USA}

\author{G. Carosi}
\author{N. Du}
\author{N. Robertson}
  \affiliation{Lawrence Livermore National Laboratory, Livermore, 
California 94550, USA}

\author{L. D. Duffy}
  \affiliation{Los Alamos National Laboratory, Los Alamos, New Mexico 
87545, USA}

\author{C. Boutan}
\author{N. S.~Oblath}
\author{M. S. Taubman}
\author{J.~Yang}
\author{E. Lentz}
  \affiliation{Pacific Northwest National Laboratory, Richland, Washington 
99354, USA}

\author{E. J. Daw}
\author{M. G. Perry}
  \affiliation{University of Sheffield, Sheffield S3 7RH, UK}

\author{C. Bartram}
  \affiliation{SLAC National Accelerator Laboratory, Menlo Park, 
California 
94025, USA}

\author{J. H. Buckley}
\author{C. Gaikwad}
\author{J. Hoffman}
\author{K. W. Murch}
  \affiliation{Washington University, St. Louis, Missouri 63130, USA}

\author{T. Nitta}
  \affiliation{International Center for Elementary Particle Physics,
The University of Tokyo, Tokyo 113-0033, Japan}

\vskip 1cm

\begin{abstract}

We investigate reentrant and dielectric loaded cavities for 
the purpose of extending the range of axion cavity haloscopes 
to lower masses, below the range where the Axion Dark Matter 
eXperiment (ADMX) has already searched. Reentrant and dielectric 
loaded cavities were simulated numerically to calculate and 
optimize their form factors and quality factors. A prototype 
reentrant cavity was built and its measured properties were 
compared with the simulations.  We estimate the sensitivity 
of axion dark matter searches using reentrant and dielectric 
loaded cavities inserted in the existing ADMX magnet at the 
University of Washington and a large magnet being installed 
at Fermilab.     

\end{abstract}

\maketitle

\section{Introduction}

Observations imply that a large fraction, of order 27\%, of 
the energy density of the Universe is some unknown substance
called "dark matter"\cite{Bertone}.   Although we do not 
know the identity of the particles constituting dark matter, 
we know two essential properties: the dark matter particles
must be collisionless and cold.  {\it Cold} means that 
their primordial velocity dispersion is sufficiently small, 
less than about $10^{-8}~c$ today, so that it may be set equal 
to zero as far as the formation of large scale structure 
and galactic halos is concerned. {\it Collisionless} means 
that the dark matter particles have, in first approximation, 
only gravitational interactions. Particles with the required 
properties are referred to as ‘cold dark matter’ (CDM).
Axions produced in the early universe by the process of 
``vacuum realignment' \cite{axdm} have those properties
\cite{Ipser}.  Axions were originally proposed to solve 
the strong CP problem of the Standard Model of elementary 
particles \cite{PQ,WW}, i.e. the puzzle of why the strong 
interactions conserve the discrete symmetries of parity 
P and charge conjugation times parity CP even though the 
Standard Model as a whole violates those symmetries.  
Therefore, the axion has a double motivation: it solves 
the strong CP and the dark matter problems simultaneously.

The axion is the quasi-Nambu-Goldstone boson associated with 
the spontaneous breaking of the U$_{\rm PQ}$(1) quasi-symmetry 
that Peccei and Quinn postulated to solve the strong CP 
problem.  Its properties depend mainly on a single unknown 
parameter $f_a$, called the axion decay constant, of order 
the energy scale at which U$_{\rm PQ}$(1) is spontaneously 
broken \cite{KSVZ,DFSZ}. The axion mass and all axion 
couplings are inversely proportional to $f_a$.  The
axion mass is given by 
\begin{equation}
m_a \simeq 6 \cdot 10^{-6}~{\rm eV}
\left({10^{12}~{\rm GeV} \over f_a}\right)~~\ .
\label{mass}
\end{equation}
The cold axion cosmological energy density is a decreasing
function of increasing axion mass \cite{axdm}.  The axion 
mass for which the cold axion cosmological energy density 
equals that of dark matter is of order $10^{-5}$ eV, 
corresponding to $f_a$ of order $10^{12}$ GeV, however
with very large uncertainties \cite{axrev}.  A major 
source of uncertainty is whether cosmological inflation 
happens before or after the phase transition in which 
U$_{\rm PQ}$(1) is spontaneously broken.  In the 
post-inflationary scenario, i.e. if the PQ phase 
transition occurs after inflation, the cold axion 
cosmological energy density has contributions from axion 
string and axion wall decay in addition to the well-known 
contribution from vacuum realignment.  Axion masses larger 
than $10^{-5}$ eV are favored in the post-inflationary 
scenario because of these extra contributions.

In the pre-inflationary scenario, i.e. if the PQ phase 
transition occurs before inflation, the cold axion 
cosmological energy density receives a contribution 
from vacuum realignment only.   The vacuum realignment
contribution is uncertain in the pre-inflationary scenario 
because it depends on the misalignment angle $\alpha_1$ 
of the axion field with respect to the minimum of its 
effective potential, at the start of the QCD phase 
transition.  As a fraction of the critical energy 
density for closing the universe, the cold axion 
energy density in the pre-inflationary scenario is 
of order \cite{axrev}
\begin{equation}
\Omega_a \sim 0.15 
\left({f_a \over 10^{12}~{\rm GeV}}\right)^{7 \over 6}
\alpha_1^2 ~~\ .
\label{caced}
\end{equation}
Since $\alpha_1$ is presumably a random number 
between $-\pi$ and $+\pi$, having the same value 
everywhere in the visible universe because the 
axion field was homogenized during inflation
\cite{SYPi}, there is 10\% probability that 
$\Omega_a$ is reduced by a factor 100, a 1\% 
probability that it is reduced by a factor 
$10^{4}$, and so on. For this reason, axion 
masses less than $10^{-5}$ eV are favored in 
the pre-inflationary scenario.

The PQ phase transition occurs at a temperature of 
order the axion decay constant $f_a \sim 10^{12}$ GeV. 
The reheat temperature at the end of inflation cannot 
be larger than the scale of inflation $\Lambda_I$ and
in many models it is of order or less than the Hubble 
rate $H_I$ during inflation \cite{Weinberg}.  The 
non-observation of tensor perturbations requires 
$H_I \lesssim 6 \cdot 10^{13}$ GeV \cite{Planck18}
or, equivalently, $\Lambda_I < 1.6 \cdot 10^{16}$ GeV. 
So, there is limited room in $(H_I , f_a)$ parameter space 
for the post-infationary scenario.  On the other hand 
nothing forbids $H_I$ and the reheat temperature from 
being very low compared to $f_a$.  So there appears 
to be a lot of parameter space for the pre-infationary 
scenario, suggesting that searches for axions with 
lower masses are well-motivated.

A number of methods have been proposed to search 
for axion dark matter by direct detection on Earth.
The topic is reviewed in Ref.~\cite{RMP}.  The cavity 
haloscope method \cite{axdet}, the earliest method 
proposed, has obtained the best results so far.  A 
cavity haloscope is an electromagnetic cavity permeated 
by a strong magnetic field and instrumented to detect 
small amounts of power inside the cavity.  When one 
of the cavity's resonant frequencies equals the axion 
mass in natural units, i.e. when $\hbar \omega = m_a c^2$ 
where $\omega$ is the angular frequency of the resonant 
mode, a small amount of microwave power is deposited 
inside the cavity as a result of axion to photon 
conversion in the applied magnetic field.  The 
relevant interaction is 
\begin{equation}
{\cal L}_{a\gamma\gamma} = - g_\gamma {\alpha\over \pi}
{1 \over f_a} a(x) \vec{E}(x)\cdot\vec{B}(x)~~\ ,
\label{agamgam}
\end{equation}
where $\alpha$ is the fine structure constant, $a(x)$, 
$\vec{E}(x)$ and $\vec{B}(x)$ are respectively the 
axion, electric and magnetic fields, and $g_\gamma$ 
is a model-dependent factor of order one. In the 
Kim-Shifman-Vainsthein-Zakharov (KSVZ) model \cite{KSVZ} 
$g_\gamma \simeq - 1.01$ , whereas  $g_\gamma \simeq 0.32$ 
in the Dine-Fischler-Srednicki-Zhitnitskii (DFSZ) model 
\cite{DFSZ}. Figure 1 shows the constraints that have been 
obtained so far on the electromagnetic coupling 
strength appearing in Eq.~(\ref{agamgam}),  
$g_{a \gamma\gamma} \equiv g_\gamma {\alpha\over \pi}
{1 \over f_a}$, as a function of the axion mass.  Using 
the cavity haloscope method, the Axion Dark Matter eXperiment 
(ADMX) has achieved sufficient sensitivity to find dark mattter 
axions at the expected halo density and with DFSZ coupling, the 
weakest coupling of the two benchmark models mentioned \cite{Du, 
Braine,Bartram}.  So far, ADMX has searched in the 600 to 1050 MHz 
range. It plans to search at higher frequencies using multi-cavity 
arrays \cite{Jihee}.  In this paper we explore methods to search 
lower frequencies, from approximately 100 to 600 MHz. 

One approach to lower frequency axion dark matter detection 
is to replace the cavity with an LC circuit \cite{SST,Kahn}.  
As with a haloscope cavity, the LC circuit is placed in a magnetic 
field.  When its resonant frequency $\omega = 1/\sqrt{LC}$ equals 
the axion mass in natural units, the axion dark matter field 
drives a tiny current in the circuit. Haloscopes with LC circuits, 
instrumented with the highest signal to noise detectors available 
and placed in strong spatially extended magnetic fields, are 
expected to achieve the sensitivity required to detect axion 
dark matter on Earth. Several pilot projects have been carried 
out \cite{ABRA,SLIC,ABRA2} and larger detectors are being 
planned \cite{DMRadio}.  An LC circuit is conveniently tuned 
by a variable capacitance.  However the capacitance cannot be 
made arbitrarily small because a circuit always has some 
parasitic capacitance.  Parasitic capacitance limits the 
highest frequency at which a large LC circuit may resonate 
\cite{SST}.  For this reason, LC circuits large enough to 
be sensitive to axion dark matter are not expected to 
reach the 200 - 600 MHz frequency range which is our main 
focus.

Instead, the approach we investigate here is to lower the 
resonant frequency of the cavity haloscope.  This can be 
done of course by scaling up the physical size of the cavity, 
which indeed works very well assuming the magnetic field region 
is enlarged accordingly. However there are obvious limits as 
to how large the cavity and magnetic field region can be 
made.  We therefore explore methods to lower the resonant 
frequency of the cavity without enlarging it. One approach 
is to fill the cavity with dielectric material, causing its 
resonant frequency to decrease as 1/$\sqrt{\epsilon}$ where 
$\epsilon$ is the dielectric constant.  The form factor $C$, 
which gives the strength of the coupling of the cavity mode to 
the axion dark matter field and which is defined in Eq.~(\ref{formf}) 
below, decreases unfortunately as ${1 \over \epsilon}$ so that 
$C(f) \propto f^2$ where $f$ is frequency. A second approach 
is to modify the cavity so as to make it ``reentrant". Reentrant
cavities are described in Section IIIA. Reentrant cavities for 
axion dark matter detection were proposed and dicussed in Refs. 
\cite{UWA1,UWA2}, including the building and characterization 
of a prototype cavity. The form factor of a reentrant cavity
also decreases qualitatively as $C(f) \propto f^2$.  So it is 
not clear at the outset whether reentrant or dielectric loaded 
cavities are the better approach.  The goal of this paper is 
to explore the two approaches systematically.  In particular 
we consider several possible reentant cavity designs and 
optimize them to achieve the highest possible form and 
quality factors.  Whether one approach is better than the 
other ultimately depends on considerations such as the cost 
of low-loss dielectric materials and the lowest frequency 
that one wishes to attain with a given cavity.

The outline of our paper is as follows.  In Section II, 
we give a general description of cavity haloscopes to 
prepare for the discussions that follow.  In Section III, 
we report on numerical simulations of reentrant cavities.
We consider several designs and optimize them for axion 
dark matter detection.  In Section IV, we report on the 
measured properties of a prototype reentrant cavity that
we built to validate the results of the numerical simulations.  
In Section V we report on numerical simulations of dielectric 
loaded cavities, evaluating their form and quality factors.   
In Section VI, we estimate the sensitivity that can be 
achieved using dielectric loaded and reentrant cavities 
inserted in the existing ADMX magnet at the University of 
Washington in Seattle and in a larger magnet that will be 
installed at Fermilab.  Section VII summarizes our conclusions.
 
\section{Cavity haloscopes}

If axions constitute the dark matter halo of the 
Milky Way Galaxy, we are immersed in a pseudo-scalar 
field, the axion field, oscillating with angular 
frequency 
\begin{equation}
\omega_a = {1 \over \hbar} E_a = 
{1 \over \hbar} (m_a c^2 + {1\over 2} m_a v^2)
= {1 \over \hbar} m_a c^2 [1 + {\cal O}(10^{-6})]~~\ ,
\label{angfreq}
\end{equation}
where $v$ is the speed of halo axions with respect to us.  
The ratio of their rest mass energy to their energy spread, 
called the ``quality factor" $Q_a$ of galactic halo axions,
is expected to be of order $10^6$ because the typical speed 
of halo dark matter is $10^{-3} c$.  Note however that cold 
flows of axion dark matter are expected \cite{Ipser2}.  Such 
flows may have velocity dispersion much less than $10^{-3} c$ 
and correspondingly larger quality factors.  

Henceforth, unless stated otherwise, we adopt units in which 
$\hbar = c = k_{\rm B}$ = 1.

A cavity haloscope searches for dark matter axions on 
Earth by attempting to convert them to microwave photons 
in an electromagnetic cavity permeated by a strong magnetic 
field \cite{axdet}.  The relevant coupling is given in 
Eq.~(\ref{agamgam}). When the resonant frequency of cavity 
mode $\eta$ equals the axion mass in natural units 
($\omega_\eta = m_a$) and the quality factor $Q_a$ 
of the axion signal is large compared to the loaded 
quality factor $Q_\eta$ of the cavity in mode $\eta$,
the power deposited into the cavity by the conversion 
process is 
\begin{equation}
P_\eta =
g^2 \rho_a B_0^2 V C_\eta {1\over m_a} Q_\eta~~\ ,
\label{sigpow}
\end{equation}
where $g \equiv g_\gamma {\alpha \over \pi} {1 \over f_a}$ 
is the coupling strength appearing in Eq.~(\ref{agamgam}), 
$\rho_a$ is the energy density of dark matter axions at the
detector location, $V$ is the volume of the cavity, $B_0$ 
is a nominal magnetic field strength inside the cavity, 
and
\begin{equation}
C_\eta 
= {\Biggl( \int_V d^3 x \vec B_0 (\vec x) \cdot
\vec E_\eta (\vec x)\Biggr)^2
\over B_0^2 V~\int_V d^3 x~\epsilon (\vec x)
\vec E_\eta (\vec x)\cdot \vec E_\eta (\vec x)}~~~\ .
\label{formf}
\end{equation}
Here, $\vec{B}_0(\vec{x})$ is the actual magnetic field 
inside the cavity, $\epsilon(\vec{x})$ is the dielectric 
constant, and the $\vec{E}_\eta(\vec{x})$ is the amplitude
of the oscillating electric field in mode $\eta$.  The 
form factor $C_\eta$ expresses the coupling strength of 
mode $\eta$ to galactic halo axions.  Generally speaking, 
the cavity mode with the highest form factor is the lowest 
TM mode, with the longitudinal direction being that of the 
static magnetic field $\vec{B}_0$. 

Equation~(\ref{sigpow}) shows that the signal power is 
proportional to the loaded quality factor $Q_\eta$ of the 
cavtity when the axion signal falls exactly in the middle 
of the cavity bandwidth $B_\eta = f/Q_\eta$.  Off resonance, 
the RHS of Eq.~(\ref{sigpow}) is multiplied by the Lorentzian 
response function characteristic of driven harmonic oscillators.  
Since the Lorentzian here has width $B_\eta$, the higher the 
quality factor the more narrow the frequency range over which 
the detector is sensitive at a given time.  The overall figure 
of merit of a detector \cite{RSI,RMP} is the rate at which it 
can search in frequency space with a given signal to noise ratio 
$s/n$.  The search rate is given by 
\begin{equation}
{d f \over dt} = {1 \over (s/n)^2} 
\left({P_\eta \over T_{\rm n}}\right)^2 
~{4 \over 9}~{Q_a \over Q_\eta}~~\ ,  
\label{frate}
\end{equation}
where $T_{\rm n}$ is the system noise temperature for 
detecting the microwave photons from axion conversion.  
Equation (\ref{frate}) assumes that the cavity bandwidth 
is wider than the axion signal bandwidth ($Q_a > Q_\eta$) 
and that the loaded quality factor $Q_\eta$ equals one 
third the unloaded quality factor.   This latter condition 
maximizes the search rate for given unloaded quality factor
\cite{RSI}.  Equation (\ref{frate}) shows that the search 
rate is proportional to $C_\eta^2 Q_\eta$.  This is the 
quantity we wish to optimize.  In practice, $Q_\eta$ varies 
relatively little. So most of our focus is on optimizing 
$C_\eta$.

Both the reentrant cavity and the dielectric loaded
cavity have form factors that decrease approximately as
frequency squared as their resonant frequency $f$ is
lowered.  For this reason, it is convenient when
comparing designs to express their performance by
the variable $\alpha(f)$ defined by
\begin{equation}
C(f) = C_0 \left({f \over f_0}\right)^2 \alpha(f)~~\ ,
\label{alpha}
\end{equation}
where $f$ is the resonant frequency of the lowest TM mode,
$C(f)$ is the form factor at that frequency, and $C_0$
and $f_0$ are the form factor and resonant frequency of
the empty cavity in its TM$_{010}$ mode. For the designs
discussed here, the empty cavity is always a cylinder of
length $L$ and radius $R$, in which case $C_0 = 0.692$ and
\begin{equation}
f_0 = 574~{\rm MHz}~{0.2~{\rm m} \over R}~~\ .
\label{f0}
\end{equation}
To facilitate comparison between different designs,
$V$ in Eq.~(\ref{formf}) is always be taken to be
the volume of the empty cavity, $V = \pi R^2 L$.

\section{Numerical simulations of reentrant cavities}

Figure 2 shows a tunable reentrant cavity 
consisting of a copper cylinder and a copper movable 
post that can be pushed in and out.  The resonant 
frequency of the cavity decreases as the movable post 
is pushed in. All the reentrant cavity designs considered 
here are axially symmetric.  The axial symmetry will be 
broken by the presence of the input and output ports 
that are necessary to couple out the axion signal and 
to characterize the cavity's properties.  In this paper, 
we ignore the perturbations that the input and output 
ports introduce. 

In the limit of axial symmetry the reentrant cavity mode 
of interest for axion dark matter detection has magnetic 
and electric fields of the form:
\begin{eqnarray}
\vec{B}(\rho,z,\phi) &=& B(\rho,z) \hat{\phi}
\nonumber\\
\vec{E}(\rho,z,\phi) &=& {i \over \omega} \left[
\left({\partial B(\rho,z) \over \partial \rho} 
+ {B(\rho, z) \over \rho}\right) \hat{z} - 
{\partial B(\rho, z) \over \partial z} \hat{\rho}\right]
\label{mode}
\end{eqnarray}
where $z, \rho, \phi$ are cylindrical coordinates, and 
$\hat{z},~\hat{\rho},~\hat{\phi}$ are the corresponding 
unit vectors.  $\hat{z}$ is in the direction both of the 
axis of axial symmetry of the cavity and of the static 
magnetic field $\vec{B}_0$.  The mode with the largest 
form factor $C$ is the lowest frequency mode of the type 
given in Eqs.~(\ref{mode}).  It becomes the TM$_{010}$ 
mode of the empty cylindrical cavity, when the post is 
removed. 
  
In the remainder of this section, we report on reentrant 
cavity simulations aimed at identifying the design of the 
movable post(s) that yield the largest form factors and 
quality factors for axion dark matter searches.  

\subsection{Single movable post}

Figure 2 shows a reentrant cavity tuned by a simple movable 
post. The figure defines the cavity length $L$ and radius 
$R$, the radius $r$ and insertion depth $d$ of the movable 
post, and the rounding radius $a$ of the edge of the movable 
post's end. 

Figure 3 shows a map of the amplitude of the longitudinal 
component $E_z(z,\rho)$ of the electric field in the lowest 
TM mode of a reentrant cavity. The form factor $C$ is 
proportional to the square of the volume integral of 
$E_z(z,\rho)$; see Eq.~(\ref{formf}).  Figure 3 shows 
that the main contribution to $C$ comes from the region 
near the end of the post.

Figure 4 shows the resonant frequency $f$ as a function of 
the insertion depth $d$ of the movable post for the case 
$L$ = 1 m, $R$ = 0.2 m, $r$ = 0.1 m, and $a$ = 20 mm.  
The reentrant cavity in its lowest TM mode can be 
understood qualitatively as an LC circuit. The end of 
the movable post forms a capacitance with the bottom of
the cavity.  The oscillating current that charges and 
discharges this capacitance flows up (down) the post and 
down (up) the sides of the cavity.  The circuit composed of 
the post and cavity sides has inductance.  This inductance 
and the capacitance both increase as the post is inserted 
deeper, causing the resonant frequency to decrease.  The 
resonant frequencies of the cavity TE modes do not change 
much as the post in inserted. There are no TEM modes. As 
a result the lowest TM mode does not cross any other modes.  
This is an advantage of the reentrant cavity approach since 
the frequency intervals where mode crossings occur require 
special treatment in axion cavity haloscope searches.  
 
Figure 5(a) shows the form factor $C$ as a function of 
frequency $f$ for three different radii of the 
movable post.  Qualitatively $C \propto f^2$.  
Figure 5(b) displays the same information in terms of 
the factor $\alpha$ defined in Eq.~(\ref{alpha}).
The behavior shown is characteristic.  By definition, 
$\alpha$ goes to one as $f \rightarrow f_0$.  Below 
$f_0$, $\alpha$ decreases quickly to a minimum of 
order 0.25 for $f \sim 0.6 f_0$ and then increases 
again as $f$ is lowered further.  The minimum 
value depends on the length of the cavity as 
discussed below.  At the lowest frequencies, 
where the search is most challenging because 
the form factor is lowest, the best performance 
in $\alpha$ is achieved by a post whose radius 
is about half the cavity radius. 

Figure 6(a) shows the product of the quality factor 
$Q$ times the skin depth $\delta$ as a function 
of frequency for several values of the movable 
post radius $r$.  The skin depth is assumed to 
be the same on all interior surfaces of the 
cavity. The quality factor is then equal to 
the product of 1/$\delta$ times a factor which 
has dimension of lengh, is proportional to the 
overall size of the cavity and depends on its 
geometry.  It is that factor which is plotted 
in the figure.  The figure shows that the quality 
factor reaches a maximum when $r \simeq R/4$.
Figure 6(b) shows $\alpha^2 Q \delta/R$ as a 
function of frequency for several values of the 
movable post radius $r$.  $\alpha^2 Q \delta$ 
may be considered a figure of merit for a 
particular geometry since the search rate 
is proportional to $C^2 Q$; see Eq.~(\ref{frate}).  
At the lowest frequencies, where the search is most
challenging because the form factor $C$ is smallest,
$\alpha^2 Q \delta$ is largest when $r \simeq 0.45 R$.

Figure 7 shows the dependence of $\alpha$ and 
$Q\delta/R$ on the curvature radius $a$ by which 
the edge of the post endcap is rounded.  The figure, 
which have a suppressed origin, shows that there is 
very little such dependence. It appears that rounding 
off the edge of the post's endcap decreases both the 
form factor and the quality factor slightly in the 
case of a single post.

Figure 8 shows $\alpha$ and $Q \delta/R$ when the 
cavity's length $L$ is decreased, from 1.0 m to 0.55 m, 
while keeping fixed $R$ = 0.2 m, $r$ = 0.1 m and $d$ = 
0.5 m.  The figure
shows that the product $L \alpha$ is nearly constant 
over the $L$ range shown. This is consistent with the 
fact that most of the contribution to the form factor 
comes from the region near the end of the movable post.  
In Fig. 3 the bottom third of the cavity is seen to 
contribute almost nothing to the form factor.   It 
can therefore be removed without changing the sensitivity 
of the experiment. If we remove the bottom third, $L$ 
and $V$ are reduced by the factor 2/3, but $C V$ is 
approximately constant, as $C$ and $\alpha$ are 
multiplied by 3/2 approximately.  The volume removed 
at no expense in sensitivity may be used for other 
purposes such as the placement of the mechanism that 
moves the tuning post.  Alternatively, provided 
$L >  2 d$, the $\alpha$ factor may be approximately 
doubled by installing a symmetrically placed post at 
the bottom of the cavity.  In this case, to avoid
mode localization in the bottom or top half of the 
caity, the two oppositely placed posts should be 
identical and have equal insertion depths.   

In this subsection, we reported on simulations of a 
reentrant cavity of lenght $L$ = 1 m and radius 
$R$ = 0.2 m, because that is the volume available 
inside the bore of the ADMX magnet at the University 
of Washington in Seattle.  However, the results are
applicable to practically any cylindrical cavity by 
exploiting 1) the fact that the dependence of the 
dimensionless quantities $\alpha$ and $Q \delta/R$ 
on the dimensionless variables $L/R$, $r/R$ and $d/R$ 
is universal, and 2) the fact that $\alpha L$ is 
approximately constant as long $L$ as does not 
approach $d$.

We sought ways to improve the performance of the 
reentrant cavity by modifying the design of the 
movable post.  We found many modifications that 
made the performance worse.  One modification 
that improves performance is to add to the first 
post a second concentric post with smaller radius.  
This is discussed in Subsection III.B.  The performance 
can be improved still further by having a series of 
concentric posts with decreasing radii, as discussed 
in Subsection III.C.   One example of a modification 
that decreases performance is to add a disk at the 
end of the movable post, as illustrated in Fig. 9.  
In all cases simulated, the addition of a disk 
decreased the form factor, regardless of the 
thickness and radius of the disk.

\subsection{Fixed and movable posts}

Figure 10 shows a reentrant cavity with a fixed post and 
a coaxial movable post that can slide through the fixed 
post.  The figure defines the various dimensions of the 
posts, their lengths $d_f$ and $d_m$, their radii $r_f$
and $r_m$, and the radii $a_f$ and $a_m$ by which their
ends are rounded.  We simulated a cavity of length 
$L$ = 1 m and radius $R$ = 0.325 m.   Such a cavity 
would fill the central volume of the Extended Frquency 
Range (EFR) magnet that the ADMX Collaboration plans to 
install and operate at Fermilab.  The resonant frequency 
of the empty cavity in its lowest TM mode is $f_0$ = 353 MHz. 

Our first step is to optimize the post dimensions so as to 
obtain the largest possible figure of merit $\alpha^2 Q \delta$ 
at the lowest frequencies where we plan to operate, near 
100 MHz.  The optimization results do not depend sharply 
on the frequency where the optimization was done. The 
optimal values are $d_f$ = 0.176 m, $r_f$ = 0.19 m, 
$a_f$ = 70 mm, $r_m$ = 0.12 m and $a_m \simeq 0$.  The 
highest frequency that can be achieved is then 262 MHz, 
for $d_m = 0$. Higher frequencies, between 262 and 353 MHz,
would be explored by using the cavity with a single post, 
as was described in the previous subsection.   

Figure 11 shows the frequency of the optimized cavity as a 
function of the total insertion depth $d= d_f+d_m$.  We 
plot the relative frequency $f/f_0$, where $f_0$ is the 
frequency of the empty cavity, versus $d/L$ to emphasize
that the relationships between dimensionless quantities
are quasi-universal.  In the same spirit, Fig. 12 shows
the dimensionless quantities $\alpha$ and $Q \delta /R$ 
as a function of $f/f_0$ for the optimized cavity, with 
$L/R = 0.308$.

It is useful to state how these results can be used
for a cavity with a different radius, e.g. $R$ = 0.2 m.  
Whatever the value of $R$, the optimal values of the post 
dimensions are $d_f \simeq 0.54 R$, $r_f \simeq 0.58 R$, 
$a_f \simeq 0.22 R$ and $r_m \simeq 0.37 R$, provided 
$d$ does not approach $L$.  If $R$ = 0.2 m 
and $L = 0.308 R$ = 0.615 m, Figs. 11 and 12 for the 
optimized cavity apply without change.  If instead $R$ 
= 0.2 m but $L$ = 1 m, Fig. 12 applies except that the
$\alpha$ values should be rescaled using the fact that 
$\alpha L$ = constant as long as $L$ does not approach 
$d$.  The comparison shows that the reentrant cavity with 
a fixed and a movable post performs somewhat better than 
the reentrant cavity with a simple movable post, improving 
the form factor typically by 10\%.  For example, at $f$ = 
344 MHz, the reentrant cavity with $R$ = 0.2 m, $L$ = 1 m 
and a simple post of radius $r$ = 0.1 m, has $\alpha$ = 
0.235.  The same cavity at the same frequency but with 
optimized fixed and movable posts has $\alpha$ = 0.264.

\subsection{Telescopic post}

To explore how much the form factor of a reentrant 
cavity can be improved further by inserting a post 
with several segments of successively smaller radii, 
we simulated the cavity depicted in Fig. 13 for the 
case $R$ = 0.325 m and $L$ = 1.1 m.  Figure 14 shows 
the amplitude of the longitudinal component $E_z(z, \rho)$ 
of the elctric field in the lowest TM mode of such a cavity.
In the simulation, the post segment with the smallest radius 
is inserted first, followed by the other segments in order 
of increasing radii.  The parameters $r_i$ and $r_o$, 
defined in Fig. 13, and the number of segments $n$ 
were varied to optimize $\alpha$ and $Q \delta$ at
low frequencies. Figures 15 shows the resulting 
$\alpha$ factor and $Q \delta/R$ as a function of 
frequency for a cavity with $n=8$, $r_0 = 0.3$ m 
and $r_i = 0.1$ m.  The simulations show that the 
telescopic post does improve the form factor at low 
frequencies, where the search is most challenging, 
compared to the fixed and movable posts setup. For 
example at 100 MHz, the frequency for which the fixed 
and movable posts were optimized, the telescopic post 
has $\alpha = 0.65$ whereas the fixed and movable 
posts setup has $\alpha = 0.60$.  

\section{Prototype Reentrant Cavity}

A prototype reentrant cavity with fixed and movable 
posts was built and tested at the University of Florida.
Figure 16 shows a schematic drawing of the cavity. It was 
built of OFHC copper and tested at room temperature. Its 
dimensions are $L$ = 0.381 m, $R$ = 77 mm, $d_f$ = 0.102 m, 
$r_f$ = 54 mm, $a_f$ = 6 mm, $r_m$ = 27 mm and $a_m = 0$.
Figure 17 shows our measurements of its resonant frequency 
as a function of total insertion depth $d = d_f + d_m$
and the prediction from the numerical simulations.  The 
agreement here is very good.  

Figure 18 shows the quality factor as a function of 
frequency and the prediction from the simulations.  The 
measured values are 15\% to 40\% lower than predicted.  
This may be due in part to an increase in the skin depth,
compared to that of pure copper, caused by lack of 
cleanliness of the cavity's inner walls.  However, 
even when a decrease in skin depth is allowed for, 
there is less than perfect agreement between theory 
and experiment in that Fig. 18 shows a dip in the 
measured values near 430 MHz whereas there is no 
such dip in the predicted values. We established 
that the dip is due to the mixing of the mode of 
interest, the cavity lowest TM mode, with a mode 
located outside the cavity.  Indeed the dip can be 
made deeper and displaced in frequency by modifying 
the cavity environment. We believe the coupling 
between the interior and exterior of the cavity occurs 
through two small holes that were made in the endplate 
of the movable post.  The holes were intended to allow 
input and output ports located there but, when the 
measurements were made, the two small holes were 
unused and left open.  This issue will be investigated 
further.  

\section{Numerical simulations of dielectric loaded cavities} 

In this section we investigate dielectric loading as a means 
to lower the resonant frequency of a cylindrical cavity in its 
lowest TM mode, keeping the cavity radius fixed. The cavity 
is tuned by moving a metallic rod sideways, a standard method 
used in cavity haloscopes.  We assume that the dielectric 
material has properties similar to those of sapphire.  
Sapphire has a dielectric constant ranging from 9.3 to 
11.5 depending on orientation and, for the purpose of 
cavity haloscopes, negligibly small dielectric losses at 
cryogenic temperatures \cite{Krupka}.  Alumina (Al$_2$O$_3$) 
has the same chemical composition as sapphire and similar 
dielectric constant, but is more economical.  High purity 
alumina has adequate dielectric losses for the purpose 
of axion haloscopes.  In Ref. \cite{Alford} the measured
$\tan \delta$ is approximately $2 \times 10^{-5}$ at room 
temperature.  In our simulations, $\epsilon = 11.1$ and 
$\tan \delta  = 10^{-7}$.
      
Figure 19 depicts cross-sectional views of  
two possible designs. The cavity in (a) has a large 
ceramic cylinder machined to allow the movement 
of the metallic tuning rod.  The cavity in (b) has 
instead many ceramic rods of small radius. Using many 
small rods is obviously simpler and more feasible in 
terms of fabrication. 

A cavity of radius $R$ = 0.325 m and length $L$ = 1 m 
was simulated in three configurations, with $N$ = 9, 26, 
and 61 dielectric rods, as shown in Fig. 20.  
In all cases, the dielectric rods have radius 12.5 mm,  
the metallic tuning rod has radius 56 mm for $N$ = 9, 
and 65 mm for $N$ = 26 and 61.  The frequency range 
165-370 MHz is covered without gaps by the three 
configurations.  The cavity with a metallic tuning 
rod of radius 50 mm and no dielectric rods covers 
370-525 MHz.

Figure 21 shows $\alpha$ and $Q \delta/R$ as a 
function of frequency for the three dielectric 
loaded configuations. Unlike reentrant cavities, 
dielectric loaded cavities are prone to mode crossings. 
In this particular example, each configuration has one 
major mode crossing where the form factor dips.  

\section{Sensitivity estimates}

In this section, we estimate the downward reach in 
frequency of reentrant and dielectric loaded cavities 
installed in the magnet presently in use by ADMX 
at the University of Wasington (UW) in Seattle and 
in the EFR magnet that ADMX plans to install and
operate at Fermilab.   For the UW magnet, we take 
$L$ = 1 m, $R$ = 0.2 m, and $B_0$ = 7.5 T.  For 
the EFR magnet $L$ = 1 m, $R$ = 0.325 m, and 
$B_0$ = 9 T. 

We make two alternative assumptions about the 
density and energy dispersion of the local dark 
matter axion density.  Assumption A is that the 
axion energy density is $\rho_a$ = 450 MeV/cc
and $Q_a$ = $10^6$.  Assumption A is what the ADMX
Collaboration has consistently used in the past to 
derive limits on the coupling $g = g_{a\gamma\gamma}$ 
from its observations.  Assumption B is the prediction 
of the galactic halo model described in Ref. \cite{Duffy}.  
Observational evidence is claimed for that model in 
Refs. \cite{Duffy,Chak}.  The model predicts the 
existence on Earth of a cold flow, called the 
"Big Flow", with energy density estimated to be 
not less than 6 GeV/cc \cite{Chak} and velocity 
dipersion not more than 70 m/s \cite{Banik}.  This 
latter upper limit implies a quality factor for the 
axion signal from the Big Flow not less than 
$0.5 \times 10^{10}$.  For assumption B, we set 
$\rho_a$ = 6 GeV/cc and $Q_a$ = $0.5 \times 10^{10}$.  

Combining Eqs.~(\ref{sigpow}) and (\ref{frate}) we have
\begin{equation}
s/n = {2 \over 3} \left({g \over g_{\rm DFSZ}}\right)^2 
g_{\rm DFSZ}^2 ~\rho_a B_0^2 V C {1 \over m_a} {1 \over T_{\rm n}} 
\sqrt{Q_\eta Q_a \over f {d \ln f \over dt}}~~\ .
\label{son}
\end{equation}
Since we ask how low in frequency a given detector may go 
with sufficient signal to noise ratio to find or rule out 
axions under stated conditions, we are interested in the 
frequency dependence on the RHS of Eq.~(\ref{son}).  We 
assume that the search rate over a logarithmic frequency 
scale ${d \ln f \over dt}$ is fixed at some reasonable 
value, similar to what ADMX has achieved in the past.  
We assume that $\alpha = 0.5$, a reasonable value for 
searches at the lowest envisaged frequencies in view 
of our simulations. Hence 
$C(f) \simeq 0.35 (f/f_0)^2$. 
We assume that the total noise temperature $T_{\rm n}$ 
is dominated by the physical temperature of the cavity, 
in which case $T_{\rm n}$ is frequency independent.
This is discussed further in the next paragraph.  
Since $g_{\rm DFSZ} \propto m_a \propto f$, and 
$Q_\eta \propto f^{1 \over 3}$ in the anomalous 
skin depth regime appropriate for the cryogenic 
temperatures at which we plan to operate the cavity, 
we have
\begin{equation}
s/n \propto \left({g \over g_{\rm DFSZ}}\right)^2 \rho_a B_0^2 V 
\sqrt{Q_a}~f^{8 \over 3} ~~\ , 
\label{son2}
\end{equation}
where we show explicitly those factors that depend on the 
magnet used (UW or EFR) and on the assumptions on the local 
axion dark matter density (A or B).  Equation (\ref{son2}) 
shows that searches with reentrant or dielectric loaded 
cavities become sharply more challenging as the frequency 
is lowered, mainly because of the worsening form factor 
and the decreasing coupling.  It is our goal to estimate 
the lowest frequency $f_{\star}$ at which the search is 
feasible under stated conditions, with the understanding 
that the search is then comparatively easy at frequencies 
larger than of order $f_{\star}$.   

The total noise temperature $T_{\rm n}$ is the sum of 
the physical temperature $T_{\rm phys}$ of the cavity 
and the system noise temperature $T_{\rm det}$ of the 
detector of microwave photons produced by axion to 
photon conversion.  Our assumption is that $T_{\rm det}$ 
is of order $T_{\rm phys}$ or less. At present the ADMX 
detector at UW achieves $T_{\rm phys} \simeq$ 150 mK 
using a dilution refrigerator. The so-called 
``quantum limit" on the noise temperature of 
receivers is $\hbar \omega/k_{\rm B}$ = 29 mK 
$\left({f \over 600~{\rm MHz}}\right)$. John Clake
and collaborators \cite{Muck} developed quantum 
limited SQUID detectors for ADMX axion searches 
in the 600-800 MHz range \cite{Aszt,Du}.  Our 
assumption regarding $T_{\rm n}$ is satisifed 
if near quantum limited receivers are developed 
and used in the 100-600 MHz range.

When estimating the sensitivity under the various 
stated assumptions we use as a benchmark the ADMX 
search reported in Ref. \cite{Du}.  It excluded DFSZ 
coupled axions in the frequency range 645-678 MHz 
using the UW magnet under assumption A on the local 
axion density.  The cavity was tuned upwards from 
the empty cavity resonant frequency $f_0$ = 574 MHz 
using a copper tuning rod.  In that search $C \simeq$ 
0.4 and $T_{\rm n} \simeq$ 300 mK.  The signal to 
noise ratio $s/n$ was approximately 4 because of 
the desirability of having of order one candidate 
signal per cavity tune from statistical fluctuations 
in the noise and the fact that the noise is Gaussian 
distributed for the medium frequency resolution 
appropriate in a search under assumption A.
 
\subsection{Using the UW magnet}

Under assumption A (Maxwell-Boltzmann) and the 
above stated conditions, Eq.~(\ref{son2}) implies 
\begin{equation}
s/n \simeq 4 \left({g \over g_{\rm DFSZ}}\right)^2 
\left({f \over 574~{\rm MHz}}\right)^{8 \over 3} ~~\ .
\label{sonWA}
\end{equation}
Since $s/n \simeq$ 4 is necessary under assumption A, 
the search is sensitive to a coupling of the axion to 
two photons
\begin{equation}
g = g_{\rm DFSZ} \left({574~\rm{MHz} \over f}\right)^{4 \over 3}~~\ .
\label{gWA}
\end{equation}
Although sensitivity to DFSZ coupling is lost, the 
search is still sensitive to KSVZ coupling over a 
wide frequency range.  It is also sensitive to many
models of dark matter in the form of axion-like particles,
as well as to QCD axions in theories where the coupling 
to two photons is enhanced for cosmological reasons, such 
as axion initial kinetic misalignement \cite{axcog,Servant}, 
or particle physics reasons such as in the photophilic hadronic 
axion model of ref. \cite{Ring}.
 
Under assumption B (Big Flow), the RHS of Eq.~(\ref{son2}) 
is multiplied by $(6/0.45) \sqrt{5 \cdot 10^3}$
compared to the search under assumption A  in view of 
the increases in $\rho_a$ and $Q_a$.  Hence
\begin{equation}
s/n \simeq 3.8 \cdot 10^3 \left({g \over g_{\rm DFSZ}}\right)^2
\left({f \over 574~{\rm MHz}}\right)^{8 \over 3} ~~\ .
\label{sonWB}
\end{equation}
On the other hand, a high resolution search that fully 
exploits assumption B requires $s/n \simeq$ 14 because 
the noise is exponentially distributed in such a search 
and there are of order $10^5$ frequency bins per cavity 
tune in which candidate signals may occur \cite{Hires}.  
Hence the lowest frequency at which sensitivity to DFSZ 
coupling can be achieved under the stated conditions is
\begin{equation}
f_\star|_{\rm UW,B} = \left({14 \over 3.8 \cdot 10^3}\right)^{3 \over 8} 
~574~{\rm MHz} = 70~{\rm MHz}~~\ .
\label{WBf}
\end{equation}

\subsection{Using the EFR magnet}

Compared to the UW magnet $B_0^2 V$ is multipled by 
$\left({9 \over 7.5}\right)^2 \left({32.5 \over 20}\right)^2$
= 3.8.  Moreover, $f_0$ is lowered to ${20 \over 32.5}$ 574 MHz 
= 353 MHz.  Under assumption A (Maxwell-Boltzmann), Eq.~(\ref{gWA}) 
is replaced by
\begin{equation} 
g = g_{\rm DFSZ} {1 \over \sqrt{3.8}}
\left({353~\rm{MHz} \over f}\right)^{4 \over 3}~~\ .
\label{gEA}
\end{equation}
The lowest frequency at which sensitivity to DFSZ
coupling can be achieved under the stated conditions 
is then
\begin{equation}
f_\star|_{\rm EFR,A} = 214~{\rm MHz}~~\ .
\label{EAf}
\end{equation}
 
Under assumption B (Big Flow), the RHS of Eq.~(\ref{sonWB}) 
is multiplied by the factor 3.8 in addition to the replacement 
574 MHz $\rightarrow$ 353 MHz for the resonant frequency of the 
empty cavity.  This yields
\begin{equation}
f_\star|_{\rm EFR,B} = 26~{\rm MHz}~~\ .
\label{EBf}
\end{equation}

\section{Summary}

We argued that there is comparatively more room in parameter 
space for inflation to occur after the PQ phase transition 
than before that transition..  This motivates axion dark matter 
searches at frequencies lower than where ADMX has already searched.  
The resonant frequency of axion haloscopes can be lowered by 
increasing the overall size of the cavity.  Although attractive 
in several respects, this approach is limited by the size of 
available magnets in which to insert the cavity.  In this 
paper we explored instead the possibility of lowering the 
resonant frequency by making the cavity reentrant or by 
loading it with dielectric material.

We simulated reentrant and dieletric loaded cavities 
numerically to compute their form factors and quality 
factors.  We explored several designs of the movable 
post, that is inserted in a reentrant cavity to tune 
it, and optimized the post dimensions. The tuning 
posts with the highest figure of merit have radii 
that decrease with depth into the cavity.  Our results 
are presented in a way which can be easily applied to 
cylindrical cavities of arbitrary dimensions.  

We built a tunable reentrant cavity and compared 
its measured properties with the simulations. 
The agreement was excellent in the plot of 
resonant frequency versus tuning post insertion 
depth.  In the plot of the quality factor versus 
frequency, the measured values were between 15\% 
to 40\% lower than the predicted ones.  The origin 
of this discrepancy is under further investigation.   

For both the reentrant cavity and the dielectric 
loaded cavity the form factor $C$ decreases with 
decreasing frequency qualitatively as $f^2$, so
that an axion search becomes progessively more
challenging as the frequency is lowered.  We 
estimated the lowest frequency $f_*$ at which 
a search with DFSZ sensitivity can be carried
out using the magnet presently used by ADMX at 
the University of Washington and the larger EFR 
magnet that ADMX plans to operate at Fermilab. The 
estimates are given by Eqs.~(\ref{WBf}), (\ref{EAf})
and (\ref{EBf}) for two alternative assumptions 
about the local dark matter density.

\begin{acknowledgments}

This work was supported by the U.S. Department of Energy
through Grants DE-SC0022148, DE-SC0009800, DESC0009723, 
DE-SC0010296, DE-SC0010280, DE-SC0011665, DEFG02-97ER41029, 
DEFG02-96ER40956, DEAC52-07NA27344, DEC03-76SF00098, and 
DESC0017987. Fermilab is a U.S. Department of Energy, 
Office of Science, HEP User Facility. Fermilab is managed 
byFermi Research Alliance, LLC (FRA), acting under Contract
No.0 DE-AC02-07CH11359. Additional support was provided by 
the Heising-Simons Foundation and by the Lawrence Livermore 
National Laboratory and Pacific Northwest National Laboratory 
LDRD offices. LLNL Release No. LLNL-JRNL825283. UWA was funded 
by the ARC Centre of Excellence for Engineered Quantum Systems, 
CE170100009, and Dark Matter Particle Physics, CE200100008. 
Ben McAllister is funded by the Forrest Research Foundation. 
Chelsea Bartram acknowledges support from the Panofsky Fellowship 
at SLAC.

\end{acknowledgments}

\newpage

\maxdeadcycles=200

\begin{figure}
\begin{center}
\includegraphics[height=110mm]{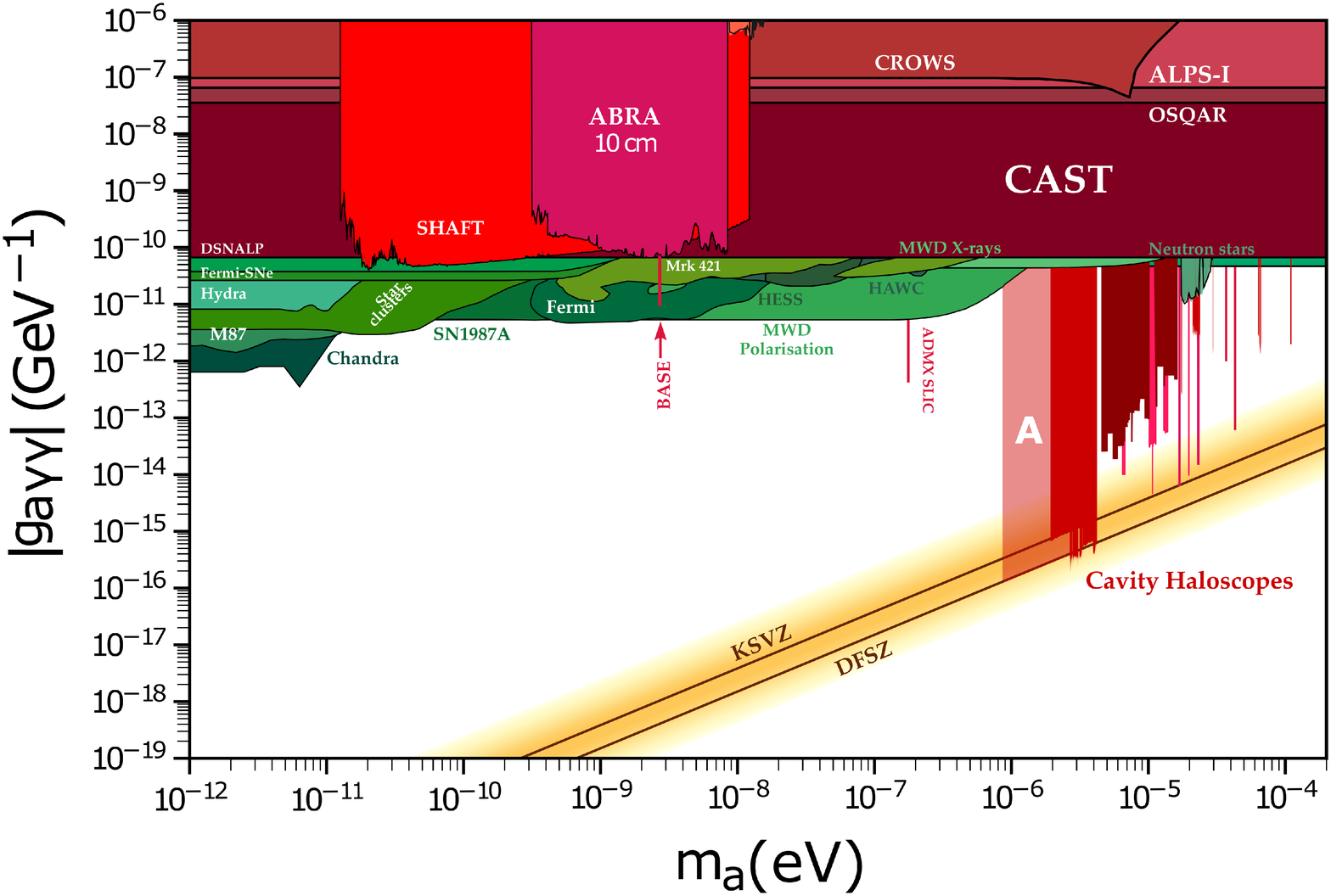}
\vspace{0.3in}
\caption{Limits on the axion electromagnetic coupling
$g_{a\gamma\gamma} \equiv g_\gamma {\alpha \over \pi}
{1 \over f_a}$ as a function of axion mass $m_a$, 
obtained by various axion dark matter searches. In 
addition, the light shaded area labeled A indicates 
the sensitivity, under assumptions spelled out in 
Section VI, of a search using a reentrant or 
dielectrically loaded cavity inserted in the 
``Extended Frequency Range" (EFR) magnet that ADMX plans 
to operate at Fermilab.  The figure was made by modifying 
Ciaran O'Hare's code 
\cite{OHare} available at https://github.com/cajohare/AxionLimits .}
\end{center}
\label{fig:axdmc}
\end{figure}

\newpage

\begin{figure}
\begin{center}
\includegraphics[height=120mm]{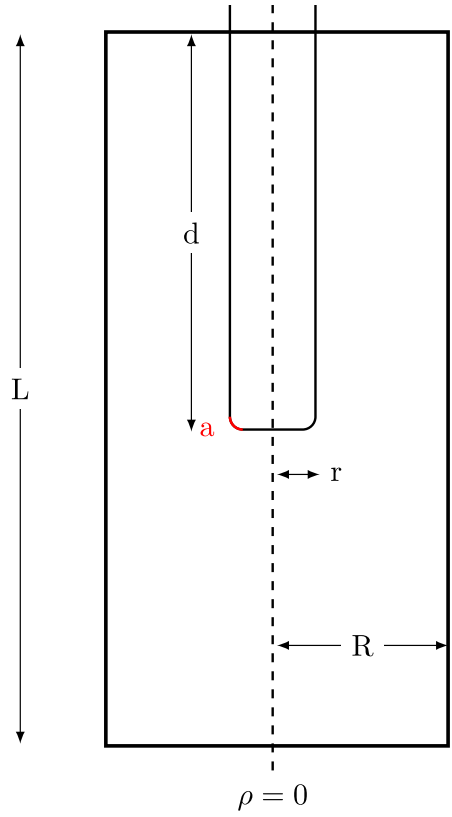}
\vspace{0.3in}
\caption{Schematic drawing of a reentrant cavity with a 
simple movable post.  The dashed line shows the cavity's 
symmetry axis. The figure defines the length $L$ and radius 
$R$ of the cavity, and the radius $r$ and insertion depth $d$ 
of the movable post.  In red is shown the curvature radius $a$ 
by which the edge of the end plate of the movable post is rounded.}
\end{center}
\label{fig:spost}
\end{figure}

\newpage

\begin{figure}
\begin{center}
\includegraphics[height=120mm]{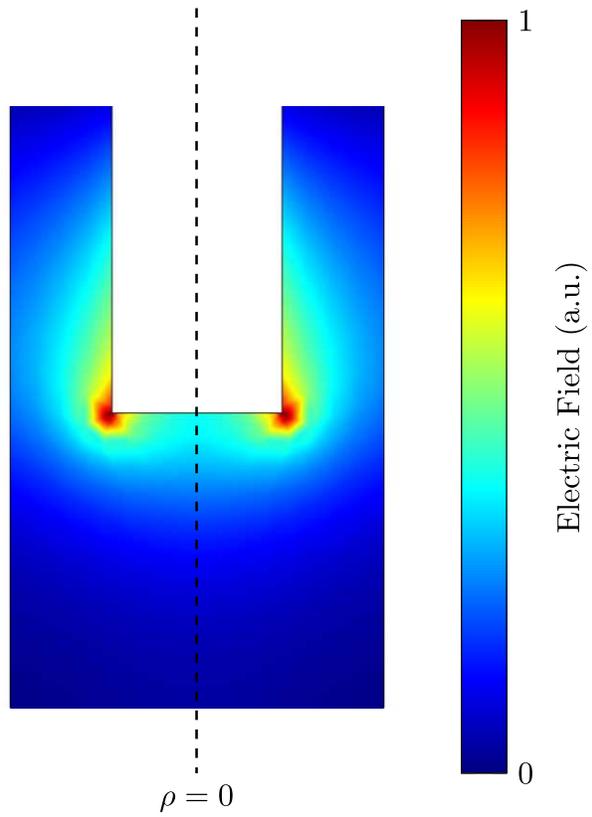}
\vspace{0.3in}
\caption{$E_z$ amplitude inside the reentrant cavity with 
a simple post for the case $L$ = 1.1 m, $R$ = 0.325 m,
$r = R/2$, $d$ = 0.55 m, and $a = 0$.}
\end{center}
\label{fig:spmode}
\end{figure}

\newpage

\begin{figure}
\begin{center}
\includegraphics[height=110mm]{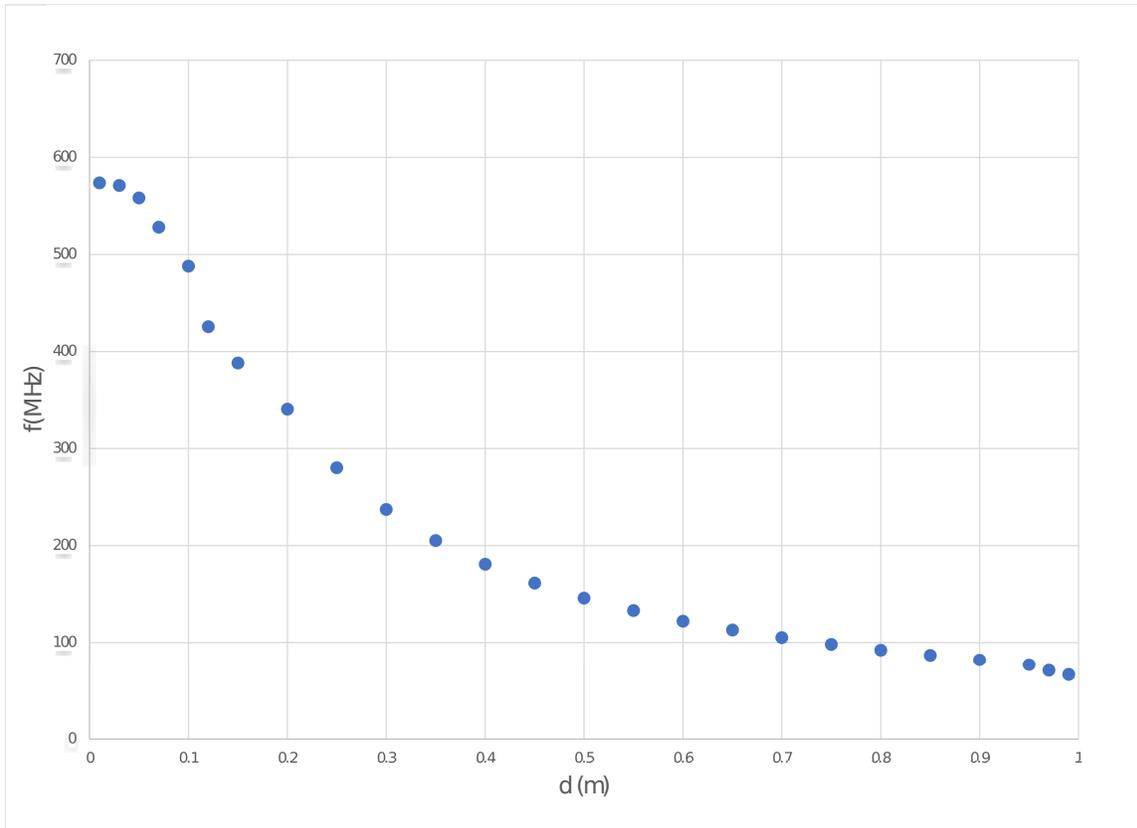}
\vspace{0.3in}
\caption{Frequency as a function of insertion depth 
of a simple post for the case $L$ = 1 m, 
$R$ = 0.2 m, $r$ = 0.1 m and $a$ = 20 mm.}
\end{center}
\label{fig:fvsd}
\end{figure}

\newpage

\begin{figure}
\begin{center}
\subfigure[]{
\includegraphics[height=95mm]{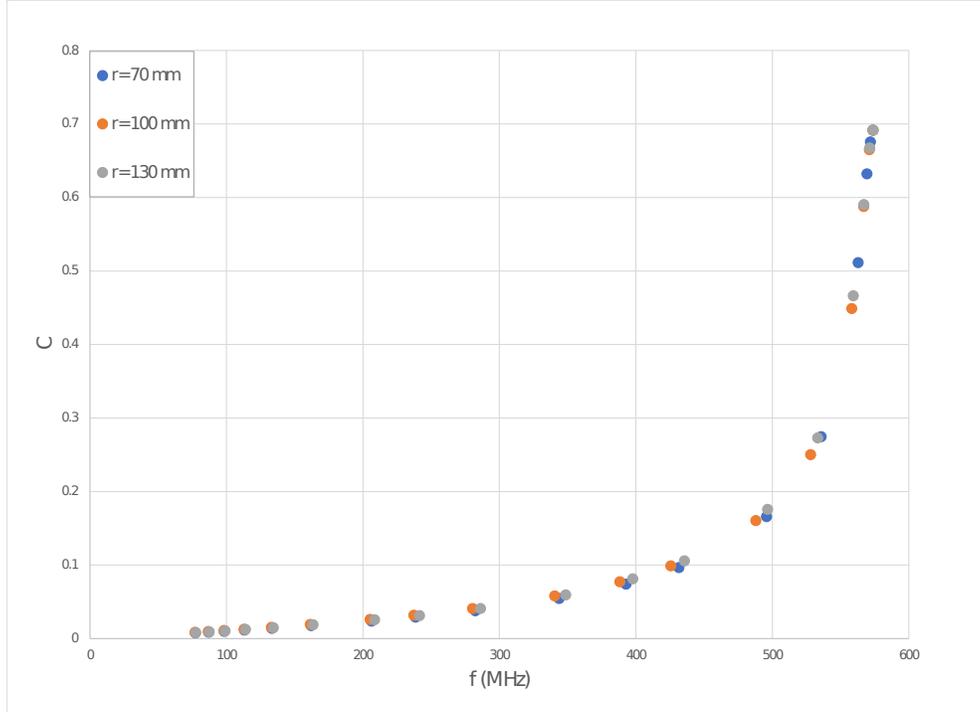}
}
\subfigure[]{
\includegraphics[height=95mm]{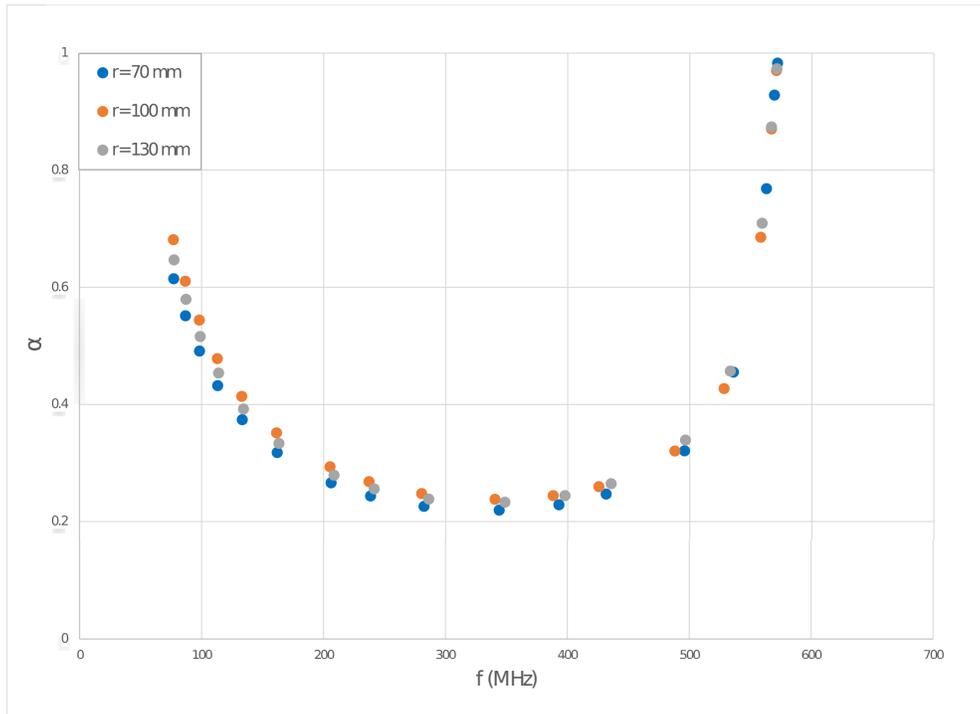}
}
\end{center}
\caption{ (a) Form factor as a function of frequency 
for three values of the movable post radius $r$,
inside a cavity of size $L$ = 1 m, $R$ = 0.2 m.  
(b) $\alpha$ as a function of frequency for the 
same data as in (a).  At low frequencies, where 
the search is most challenging, the highest form 
factors are obtained when $r$ is approximately $R/2$.}
\label{fig:Cvsf}
\end{figure}

\newpage

\begin{figure}
\begin{center}
\subfigure[]{
\includegraphics[height=95mm]{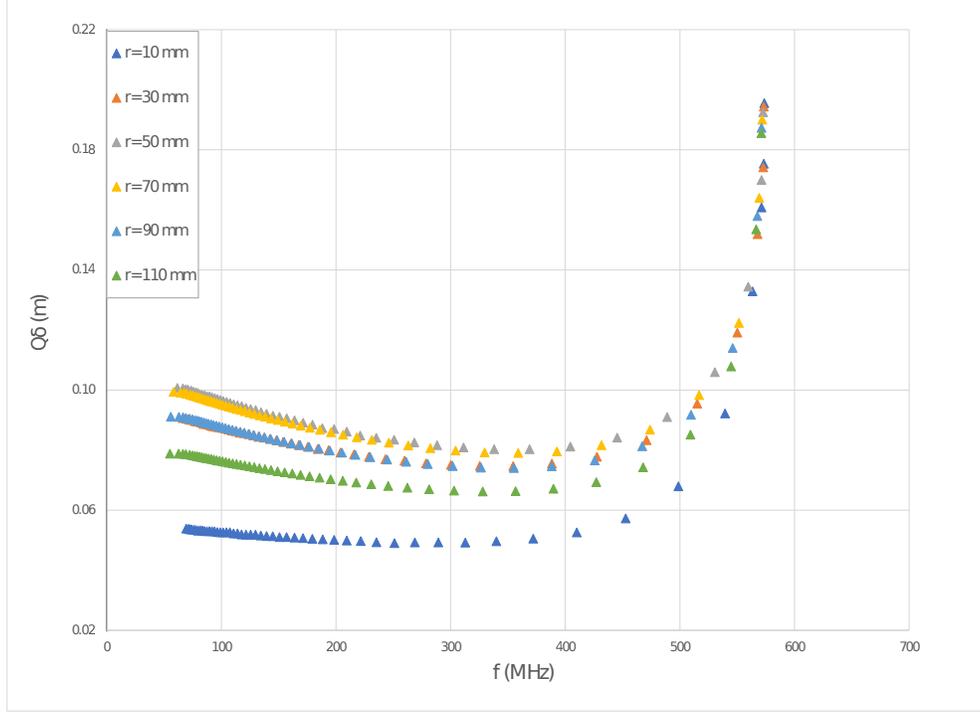}
}
\subfigure[]{
\includegraphics[height=95mm]{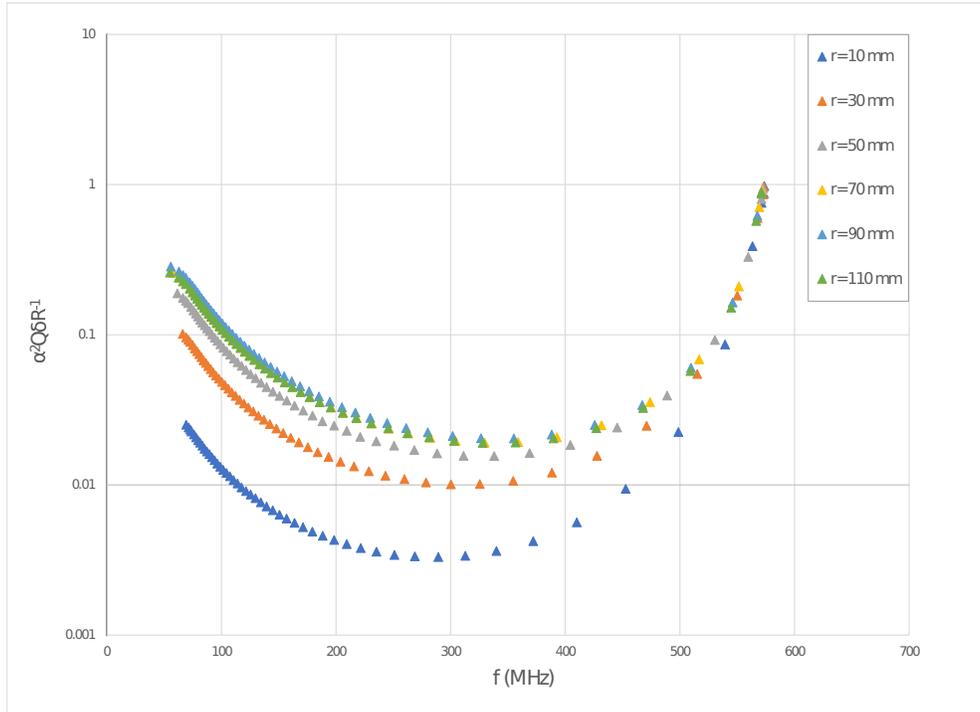}
}
\caption{(a)Product of the quality factor times the
skin depth as a function of frequency for the case
$L$ = 1 m , $R$ = 0.2 m, and several values of
the movable post radius $r$.  At low frequencies, 
where the search is most challenging, the highest 
quality factors are obtained when $r$ is approsimately 
$R/4$. (b) Figure of merit $\alpha^2 Q \delta/R$ as
a function of frequency for the case $L$ =  1 m, 
$R$ = 0.2 m, and several values of the movable post 
radius $r$,   At low frequencies, where the search 
is most challenging, the highest figure of merit 
is obtained when $r \simeq 0.45 R$.}
\end{center}
\label{fig:Qvsf}
\end{figure}

\newpage

\begin{figure}
\begin{center}
\includegraphics[height=110mm]{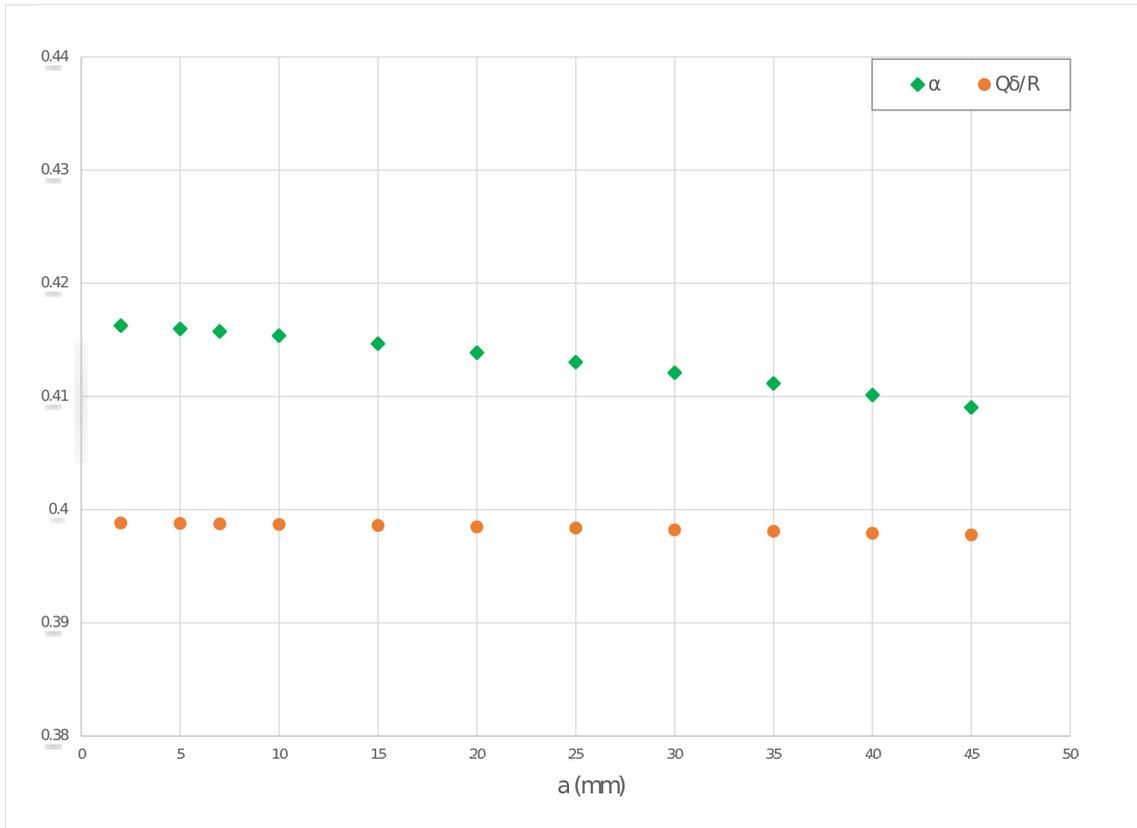}
\vspace{0.3in}
\caption{$\alpha$ and $Q \delta/R$ as a function 
of the curvature radius $a$ by which the edge of 
the movable post endcap is rounded off, for the 
case $L$ = 1 m, $R$ = 0.2 m, $r$ = 0.1 m and 
$d$ = 0.5 m.} 
\end{center}
\label{fig:alvsa}
\end{figure}

\newpage

\begin{figure}

\begin{center}
\includegraphics[height=110mm]{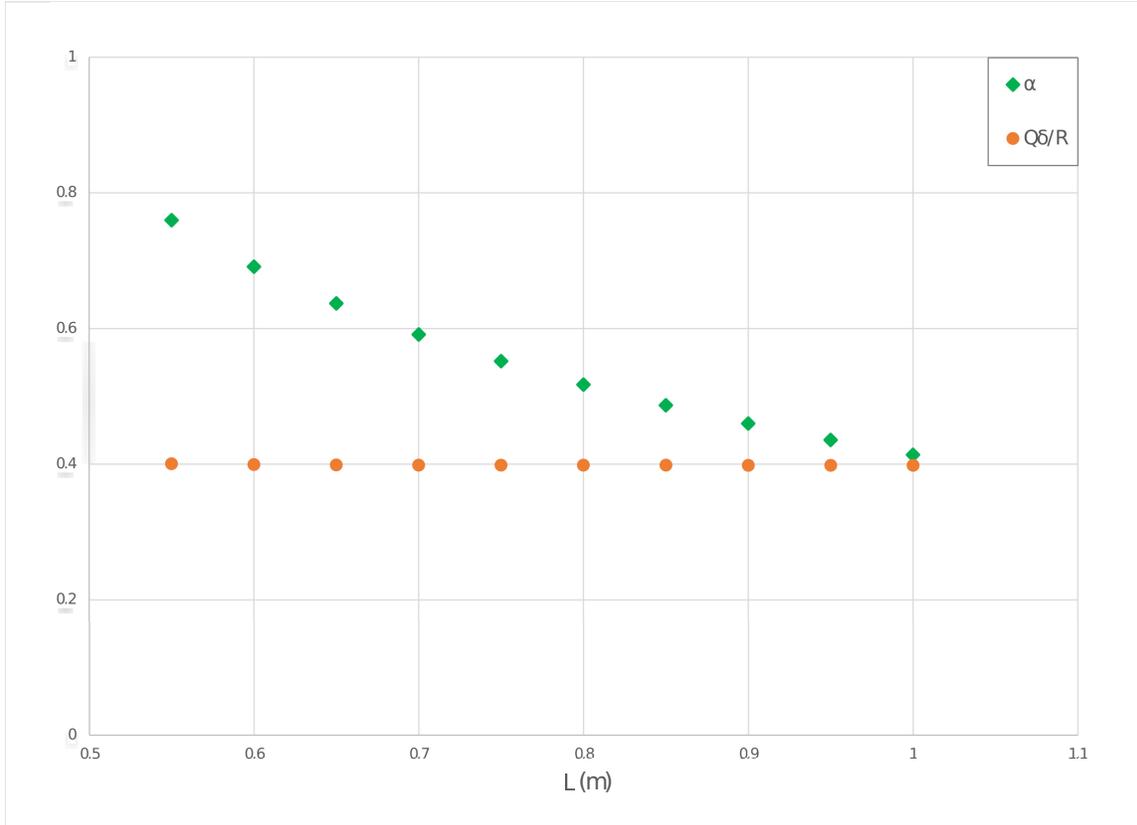}
\vspace{0.3in}
\caption{Dependence of $\alpha$ and $Q \delta/R$
on the length of the cavity $L$. Note that the 
product of $\alpha$ and $L$ and $Q \delta/R$ are
very nearly constant, consistent with the statement
that the part of the cavity away from the post
can be removed without affecting the mode of interest.}
\end{center}
\label{fig:alvsL}
\end{figure}

\newpage

\begin{figure}
\begin{center}
\includegraphics[height=120mm]{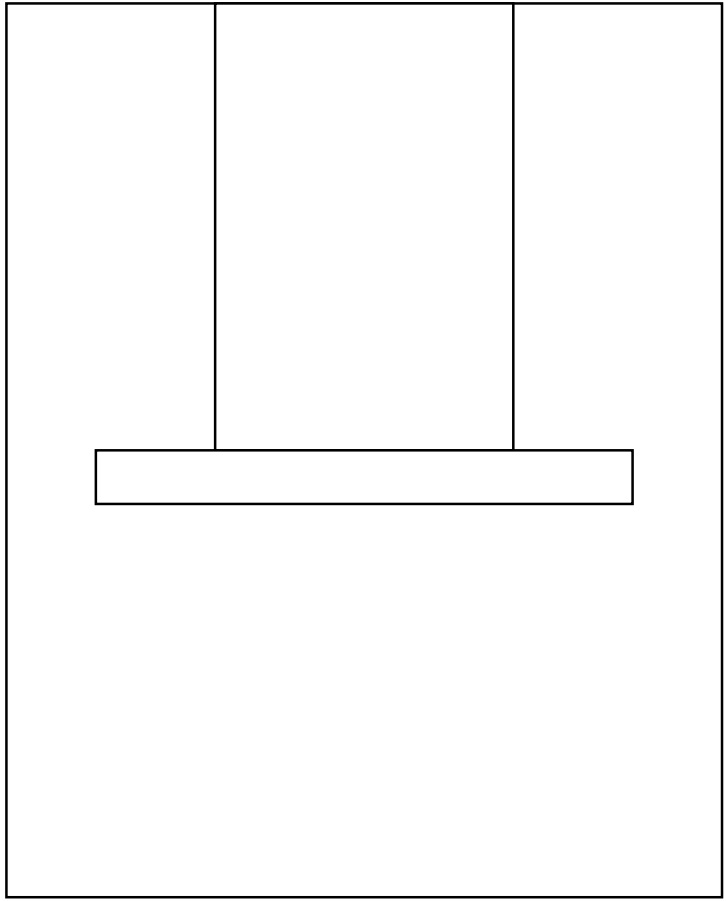}
\vspace{0.3in}
\caption{Reentrant cavity with movable post and disk.}
\end{center}
\label{fig:disk}
\end{figure}

\newpage

\begin{figure}
\begin{center}
\includegraphics[height=110mm]{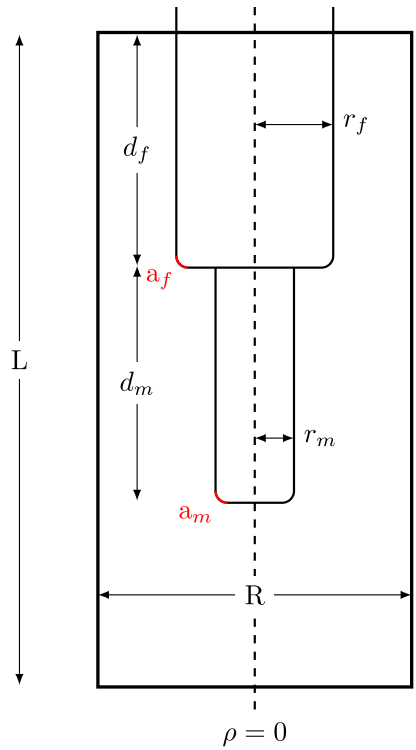}
\vspace{0.3in}
\caption{Reentrant cavity with a fixed and a movable post.
The figure defines the radii $r_f$ and $r_m$ of the fixed
and of the movable posts, their insertion depths $d_f$
and $d_m$, and the radii $a_f$ and $a_m$ by which their 
ends are rounded. The cavity is tuned by sliding the 
movable post through the fixed post.}
\end{center}
\label{fig:stub}
\end{figure}

\newpage

\begin{figure}
\begin{center}
\includegraphics[height=110mm]{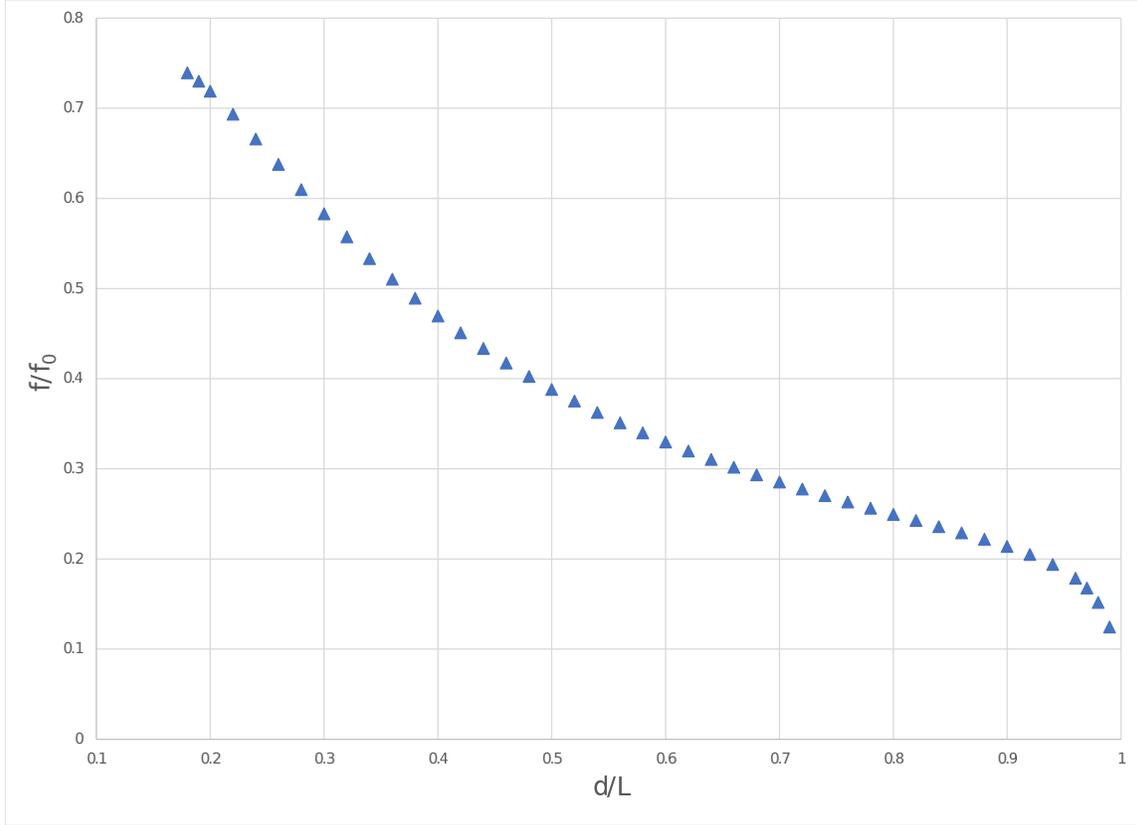}
\vspace{0.3in}
\caption{Relative frequency $f/f_0$ versus relative 
insertion depth $d/L = (d_f + d_m)/L$ for the optimized 
reentrant cavity with a fixed and a movable post. $f_0$ 
is the frequency of the empty cavity.  The cavity 
was optimized to achieve the highest figure of 
merit $\alpha^2 Q \delta/R$ at $f/f_0$ = 0.283.}
\end{center}
\label{fig:fmfvd}
\end{figure}

\newpage

\begin{figure}
\begin{center}
\includegraphics[height=110mm]{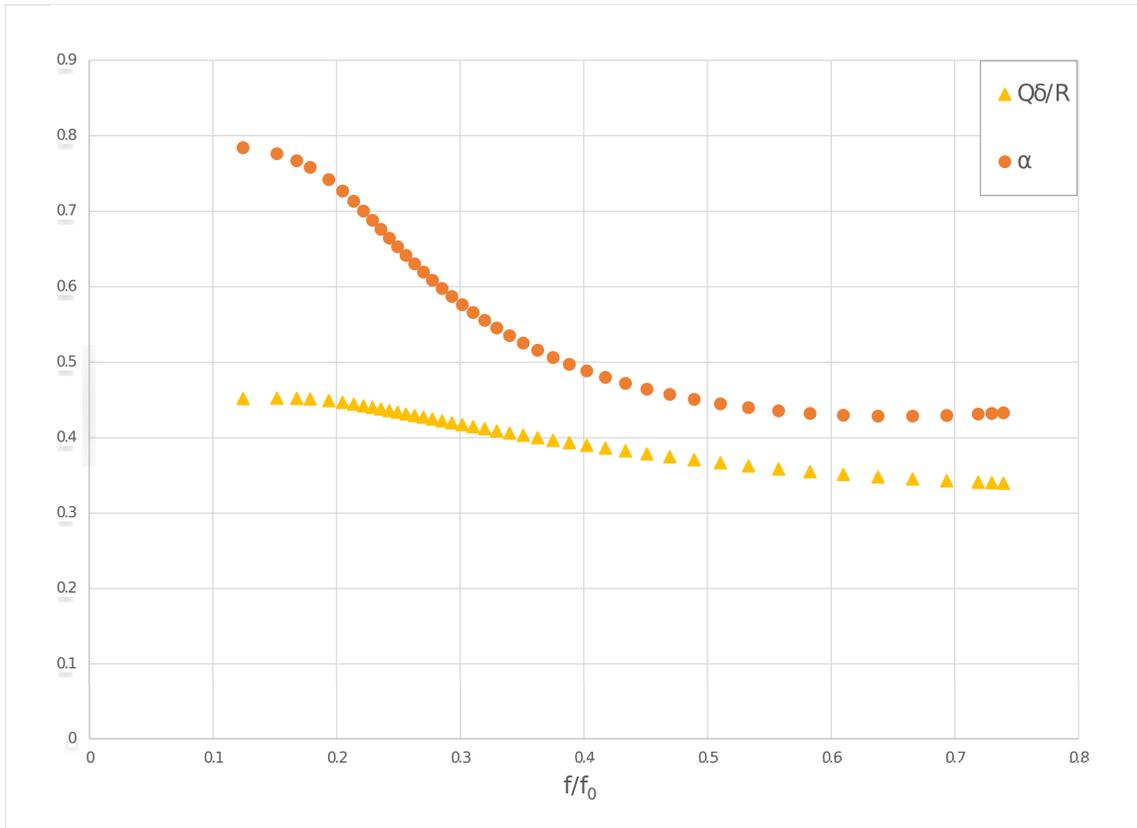}
\vspace{0.3in}
\caption{$\alpha$ and $Q \delta/R$ versus relative 
frequency for the optimized reentrant cavity with 
a fixed and a movable post.}
\end{center}
\label{fig:fmalpha}
\end{figure}

\newpage

\begin{figure}
\begin{center}
\includegraphics[height=110mm]{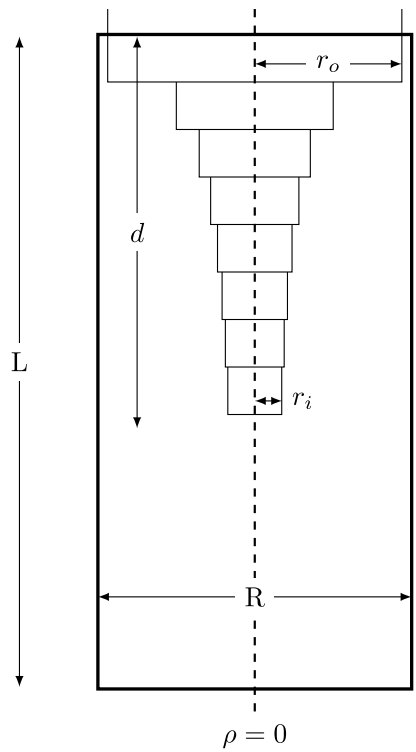}
\vspace{0.3in}
\caption{Design of simulated cavity with telescopic post.}
\end{center}
\label{fig:scope}
\end{figure}

\newpage

\begin{figure}
\begin{center}
\includegraphics[height=110mm]{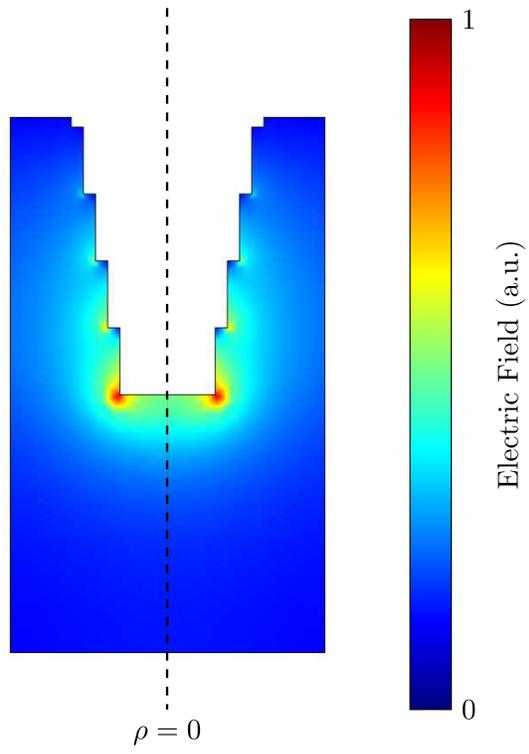}
\vspace{0.3in}
\caption{$E_z$ amplitude inside a reentrant cavity with
a telescopic post.}
\end{center}
\label{fig:scopemode}
\end{figure}

\newpage

\begin{figure}
\begin{center}
\includegraphics[height=110mm]{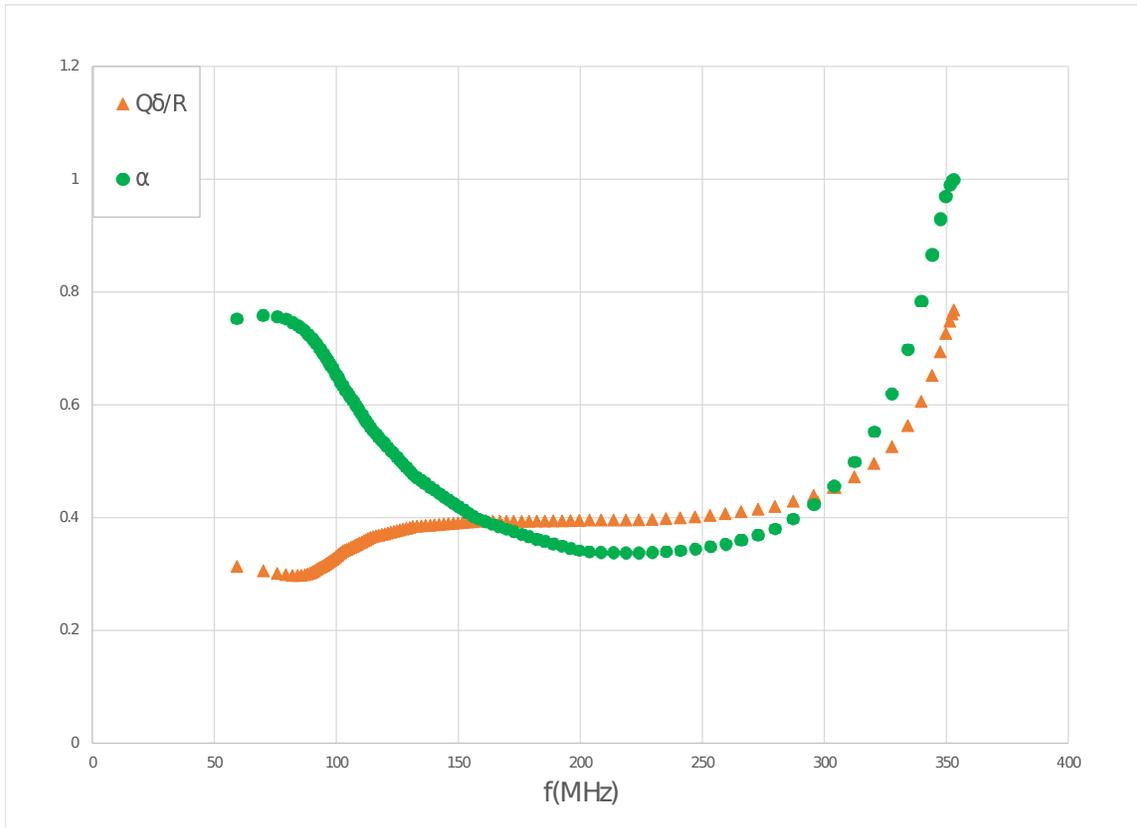}
\vspace{0.3in}
\caption{$\alpha$ and $Q \delta/R$ versus frequency 
for the cavity with telescopic post.}
\end{center}
\label{fig:TeleAlpha}
\end{figure}

\newpage

\begin{figure}
\begin{center}
\includegraphics[height=110mm]{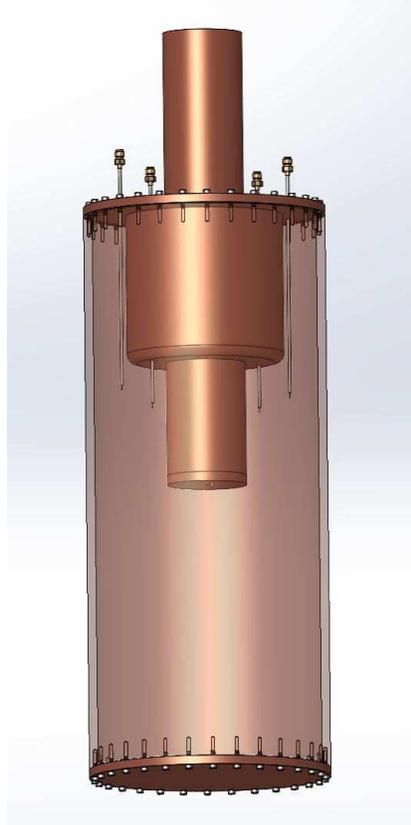}
\vspace{0.3in}
\caption{Drawing of the prototype reentrant cavity 
described in Section IV.}
\end{center}
\label{fig:prototype}
\end{figure}

\begin{figure}
\begin{center}
\includegraphics[height=110mm]{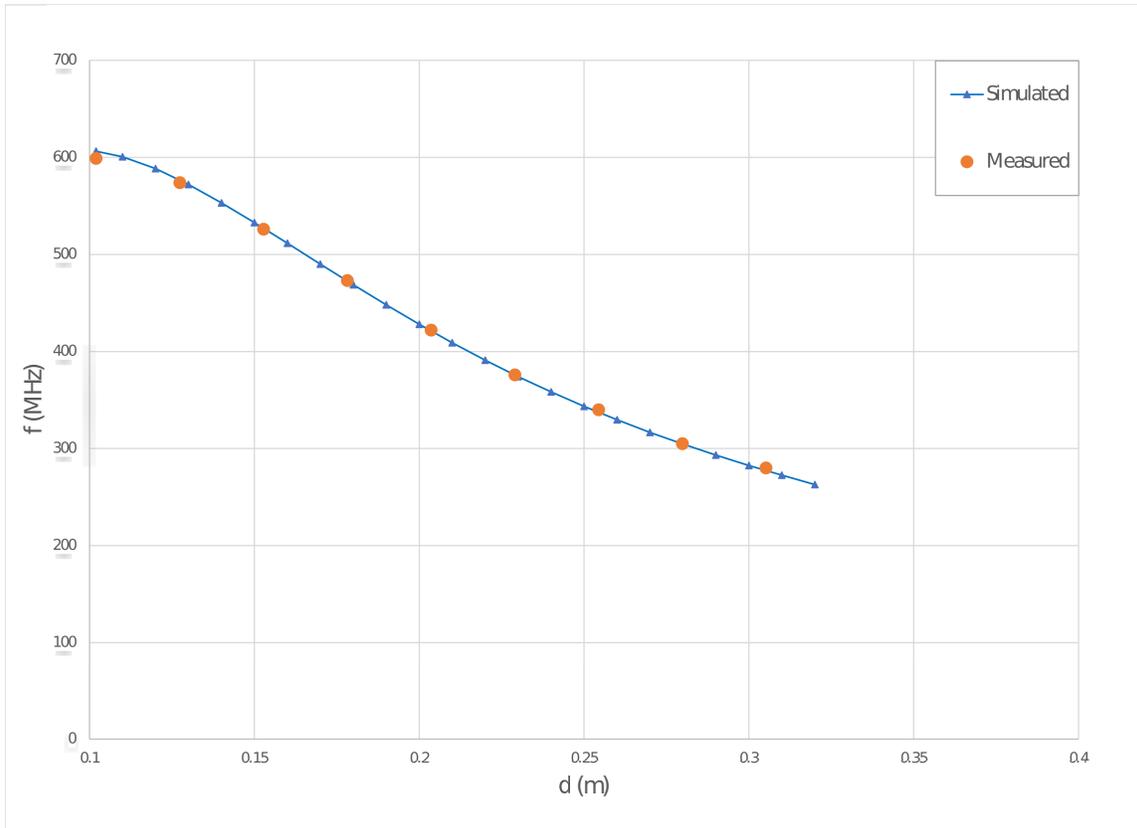}
\vspace{0.3in}
\caption{Frequency versus total 
insertion depth $d = d_f + d_m$ of the prototype 
reentrant cavity.}
\end{center}
\label{fig:prot_fvsd}
\end{figure}

\begin{figure}
\begin{center}
\includegraphics[height=110mm]{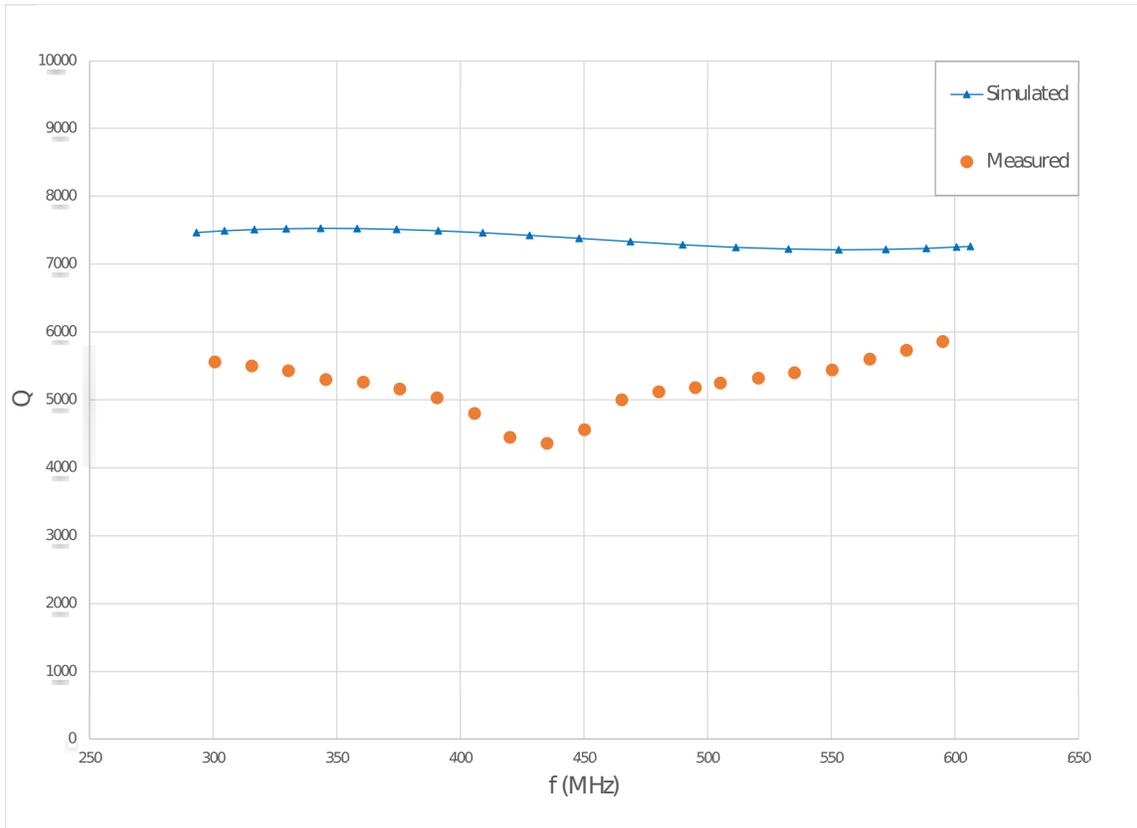}
\vspace{0.3in}
\caption{Quality factor $Q$ versus frequency of 
the prototype reentrant cavity.}
\end{center}
\label{fig:prot_Qvsf}
\end{figure}

\begin{figure} 
\begin{center} 
\includegraphics[height=80mm]{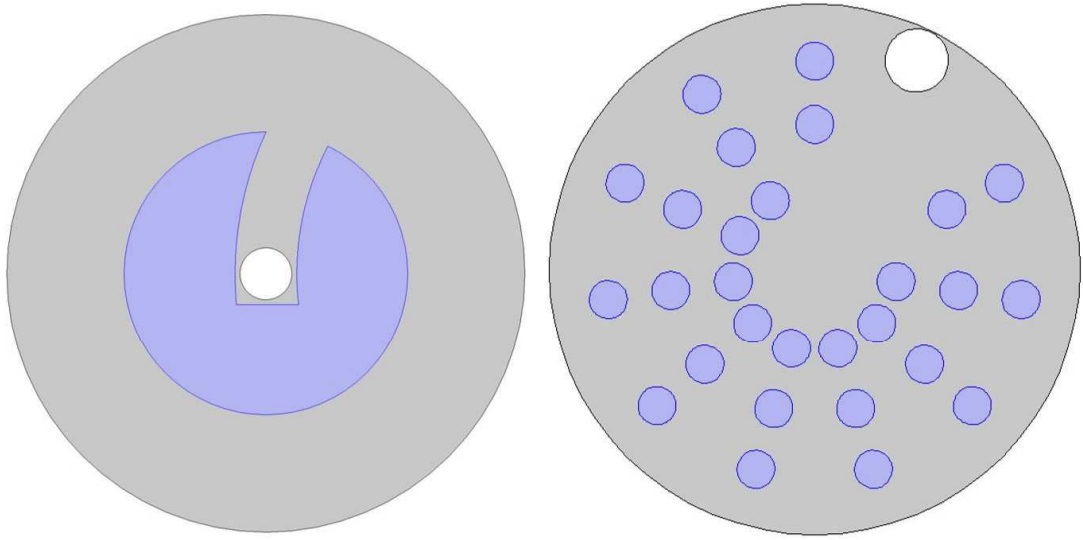} 
\vspace{0.3in} 
\caption{Cross-sectional view of dielectric loaded cavities. 
(a) Cavity loaded with a large ceramic cylinder machined to 
allow movement of a metallic tuning rod. The rod is at the 
cavity center. (b) Cavity loaded with 28 ceramic rods arranged 
to allow movement of a metallic tuning rod.  The rod is at 
the cavity wall.}
\end{center}
\label{fig:diel1}
\end{figure}

\begin{figure}
\begin{center}
\includegraphics[height=50mm]{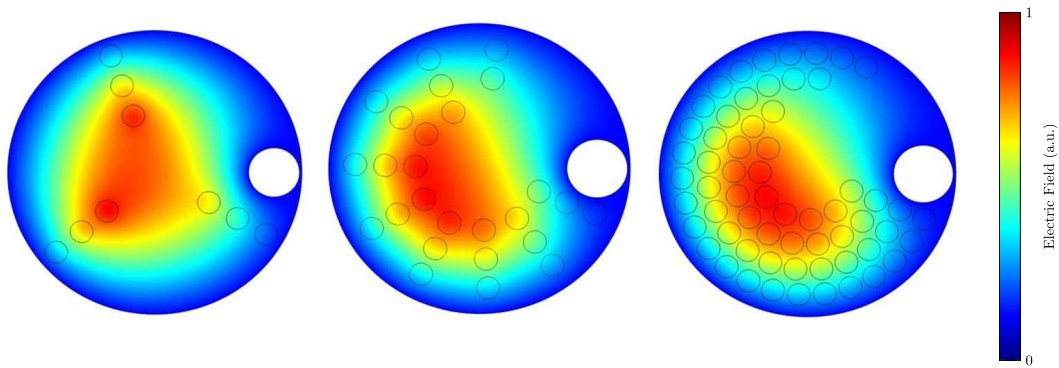}
\vspace{0.3in}
\caption{Simulation of a cavity of radius 0.325 m loaded 
with different numbers of sapphire dielectric rods, each 
12.5 mm in radius, and a metallic tuning rod. Dielectric 
loading enables the cavity to cover the frequency range 
165-370 MHz. (a) 286.43 MHz resonant mode with 9 rods, 
(b) 216.73 MHz resonant mode with 26 rods, and (c) 164.59 MHz 
resonant mode with 61 rods.}
\end{center}
\label{fig:diel2}
\end{figure}

\begin{figure}
\begin{center}
\subfigure[]{
\includegraphics[height=95mm]{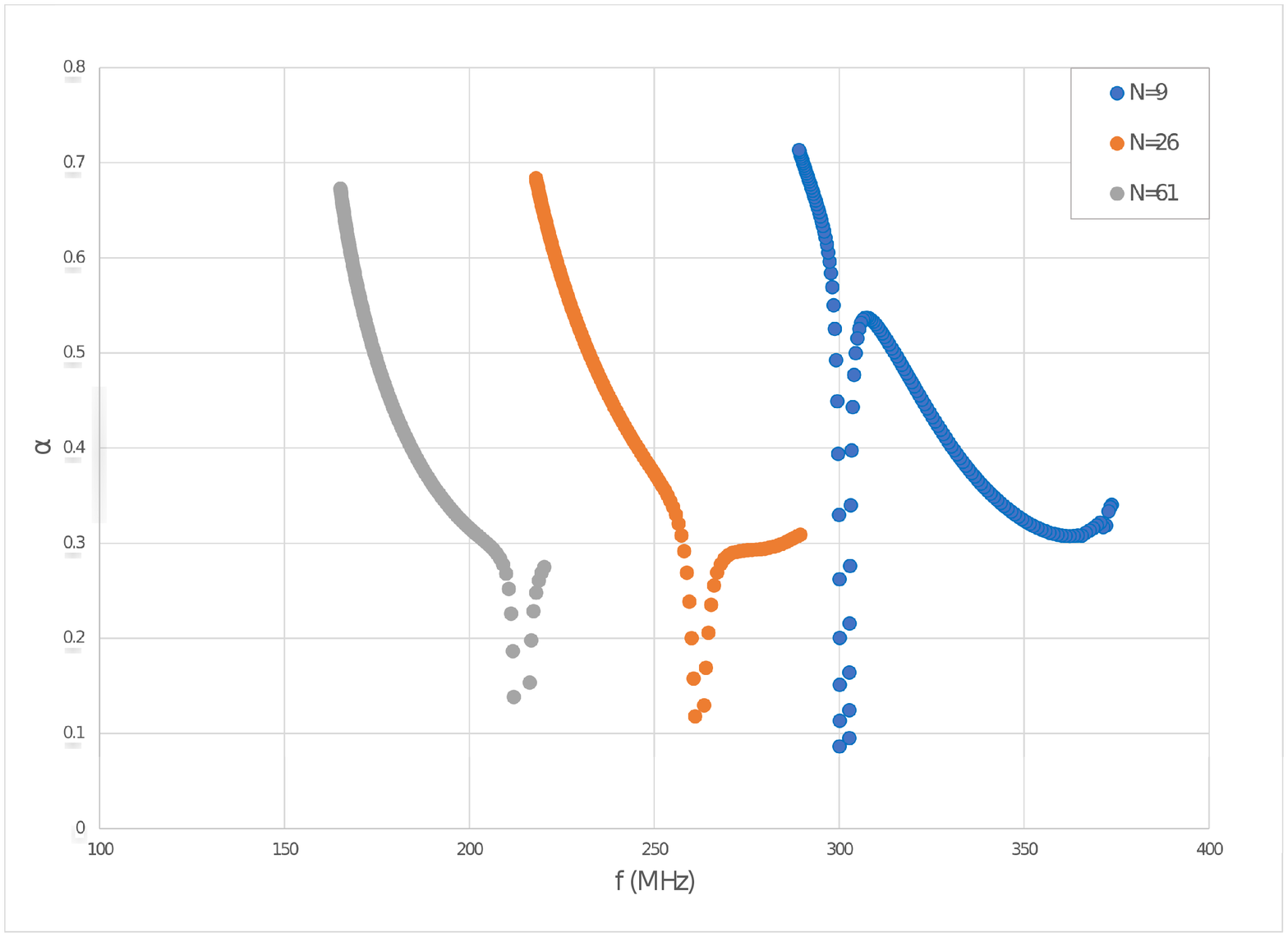}
}
\subfigure[]{
\includegraphics[height=95mm]{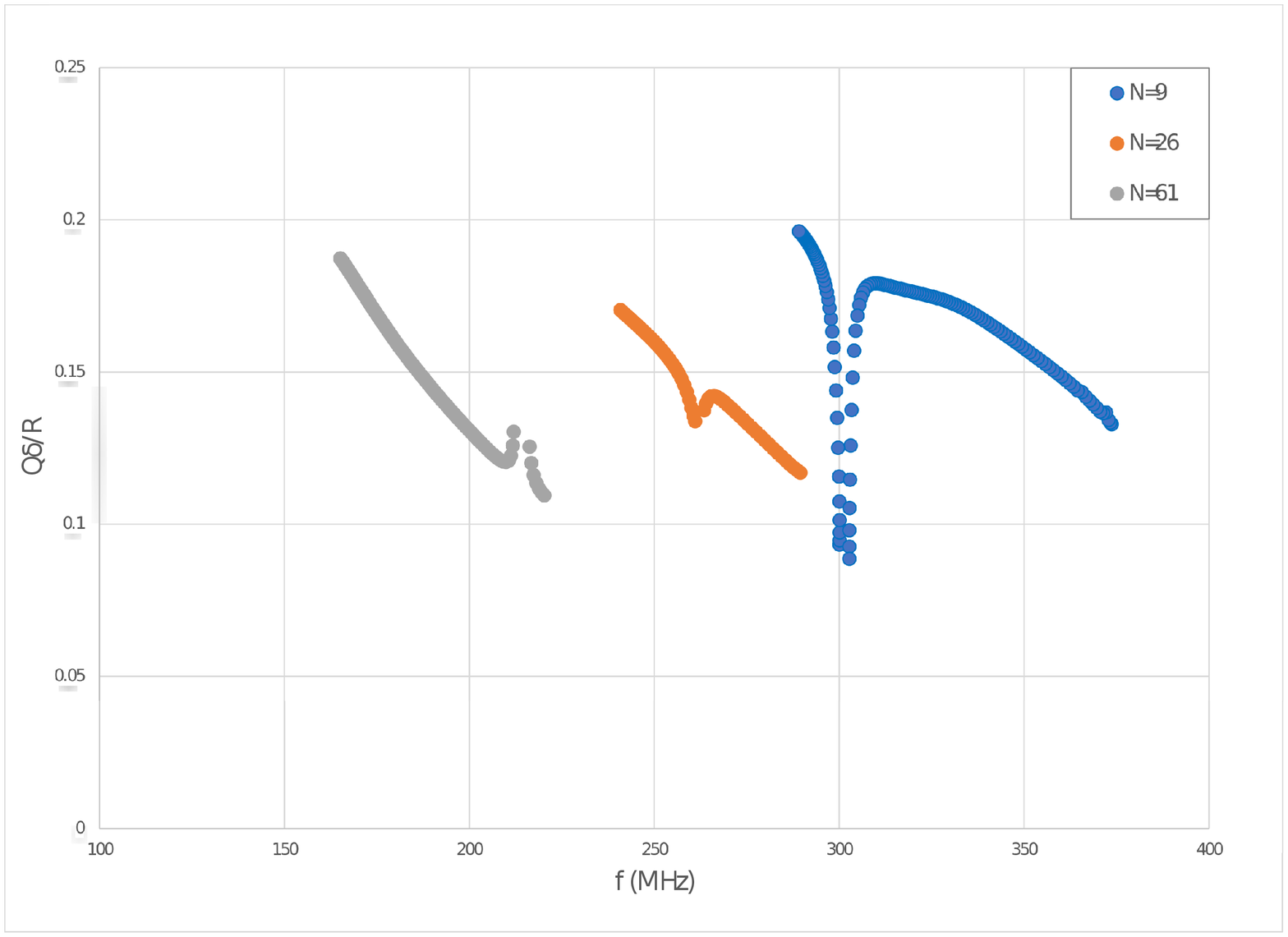}
}
\caption{(a) $\alpha$ and (b) $Q \delta/R$ 
versus frequency for the different dielectric 
loading configurations described in Fig.~20. 
The dip in each curve is due to mode crossing, 
and mixing, with a TE mode.} 
\end{center}
\label{fig:diel_avsf}
\end{figure}


\begin{thebibliography}{bib}

\bibitem{Bertone}
See, for example, {\it Particle Dark Matter}, 
edited by G. Bertone, Cambridge Uivversity Press, 2010.

\bibitem{axdm}
J. Preskill, M. Wise and F. Wilczek, Phys. Lett. B120 (1983) 127;
L. Abbott and P. Sikivie, Phys. Lett. B120 (1983) 133;
M. Dine and W. Fischler, Phys. Lett. B120 (1983) 137.

\bibitem{Ipser}
J. Ipser and P. Sikivie, Phys. Rev. Lett. 50 (1983) 925.

\bibitem{PQ}
R. D. Peccei and H. Quinn, Phys. Rev. Lett. 38 (1977) 1440
and Phys.Rev. D16 (1977) 1791.

\bibitem{WW}
S. Weinberg, Phys. Rev. Lett. 40 (1978) 223; 
F. Wilczek, Phys. Rev. Lett. 40 (1978) 279.

\bibitem{KSVZ}
J. Kim, Phys. Rev. Lett. 43 (1979) 103; M. A. Shifman,
A. I. Vainshtein and V. I. Zakharov, Nucl. Phys. B166
(1980) 493.

\bibitem{DFSZ}
M. Dine, W. Fischler and M. Srednicki, Phys. Lett. B104 
(1981) 199; A. Zhitnitskii, Sov. J. Nucl. 31 (1980) 260.

\bibitem{axrev}
Reviews of axion cosmology include:  
P. Sikivie, Lect. Notes Phys. 741 (2005) 083513;
D.J.E. Marsh, Phys. Rep. 643 (2016) 1.

\bibitem{SYPi}
S.-Y. Pi, Phys. Rev. Lett. 52 (1984) 1725.

\bibitem{Weinberg}
S. Weinberg, {\it Cosmology}, Oxford University Press, 2008.

\bibitem{Planck18}
Y. Akrami et al.(Planck Collaboration), 
Astron. and Astroph. 641 (2020) A10.

\bibitem{RMP}
P. Sikivie, Rev. Mod. Phys. 93 (2021) 015004.

\bibitem{axdet}
P. Sikivie, Phys. Rev. Lett. 51 (1983) 1415, [Erratum: Phys. Rev. Lett. 52
(1984) 695], and Phys. Rev. D32 (1985) 2988, [Erratum: Phys. Rev. D36
(1987) 974].

\bibitem{OHare}
C. O'Hare, {\it Axion Limits}, July 2022,
https://cajohare.github.io/AxionLimits/ .

\bibitem{Du}
N. Du et al., Phys. Rev. Lett. 120 (2018) 151301.

\bibitem{Braine}
T. Braine et al., Phys. Rev. Lett. 124 (2020) 101303;
C. Bartram et al., Phys. Rev. D103 (2021) 032002.

\bibitem{Bartram}
C. Bartram et al., Phys. Rev. Lett. 127 (2021) 261803.

\bibitem{Jihee}
J. Yang et al., Springer Proc. Phys. 245 (2020) 53.

\bibitem{SST}
P. Sikivie, N. Sullivan and D. Tanner, 
Phys. Rev. Lett. 112 (2014) 131301.

\bibitem{Kahn}
Y. Kahn, B.R. Safdi and J. Thaler, 
Phys. Rev. Lett. 117 (2016) 141801.

\bibitem{ABRA}
J.L. Ouellet et al., Phys. Rev. Lett. 122 (2019) 121801, 
and Phys. Rev. D99 (2019) 052012.

\bibitem{SLIC}
N. Crisosto et al., Phys. Rev. Lett. 124 (2020) 241101.

\bibitem{ABRA2}
C.P. Salemi et al., Phys. Rev. Lett. 127 (2021) 081801.

\bibitem{DMRadio}
L. Brouwer et al., arXiv:2204.13781, and 
A. AlShirawi et al., arXiv:2302.14084.

\bibitem{UWA1}
B.T. McAllister, S.R. Parker and M.E. Tobar,
Phys. Rev. D94 (2016) 042001.

\bibitem{UWA2}
B.T. McAllister et al., J. Appl. Phys. 122 (2017) 144501

\bibitem{Ipser2}
P. Sikivie and J.R. Ipser, Phys, Lett. B291 (1992) 288.

\bibitem{RSI}
C. Hagmann, P. Sikivie, N. Sullivan, D.B. Tanner and 
S.-I. Cho, Rev. Sci. Ins. 61 (1990) 1076.

\bibitem{Krupka}
J. Krupka et al., Meas. Sci. Technol. 10 (1999) 387.

\bibitem{Alford}
N.M. Alford and S.J. Penn, J. Appl. Phys. 80 (1996) 5895.

\bibitem{Duffy}
L.D. Duffy and P. Sikivie, Phys. Rev. D78 (2008) 063508.

\bibitem{Chak}
S.S. Chakrabarty, Y. Han, A.H. Gonzalez and P. Sikivie,
Phys. Dark Univ. 33(2021) 100838.

\bibitem{Banik}
N. Banik and P. Sikivie, Phys. Rev. D93 (2016) 103509.

\bibitem{Muck}
M. M\"uck et al., Appl. Phys. Lett. 72 (1998) 2885.

\bibitem{Aszt}
S.J. Asztalos et al., Phys. Rev. Lett. 104 (2010) 041301.

\bibitem{axcog}
R.T. Co, L.J. Hall and K. Harigaya, Phys. Rev. Lett.
120 (2018) 211602, and Phys. Rev. Lett. 124 (2020) 251802;

\bibitem{Servant}
C. Eroncel et al., JCAP et al. 10 (2022) 053.

\bibitem{Ring}
A.V Sokolev and A. Ringwald, JHEP 06 (2021) 123.

\bibitem{Hires}
L.D. Duffy et al., Phys. Rev. Lett. 95 (2005) 091304;
L.D. Duffy et al., Phys. Rev. D74 (2006) 012006; 
J. Hoskins et al., Phys. Rev. D84 (2011) 121302.

\end{thebibliography}
\end{document}